\begingroup\expandafter\expandafter\expandafter\endgroup
\expandafter\ifx\csname pdfsuppresswarningpagegroup\endcsname\relax
\else
\fi
\documentclass[acmsmall,manuscript,screen,nonacm]{acmart}

\usepackage[utf8]{inputenc}

\usepackage{geometry}
\usepackage{xcolor}
\usepackage{graphicx}
\usepackage{subcaption}

\usepackage{amsmath}
\usepackage{mathtools}

\usepackage{algpseudocode}
\usepackage{algorithmicx}
\usepackage{algorithm}

\usepackage[algo2e,ruled,vlined,linesnumbered]{algorithm2e}
\SetCommentSty{scriptsize}
\SetKwComment{tcc}{[}{]}
\SetKwIF{If}{ElseIf}{Else}{if}{}{else if}{else}{endif}
\SetKwFor{While}{while}{}{endw}
\SetKwFor{ForEach}{for each}{}{endfch}
\SetKwFor{For}{for}{}{}%
\SetKwComment{tcp}{$\rightarrow$\quad}{}

\newlength{\commentWidth}
\newcommand{\atcp}[1]{\tcp*[r]{\makebox[\commentWidth]{#1\hfill}}}

\usepackage{enumitem}
\usepackage{array}
\usepackage{booktabs}
\usepackage{setspace}

\usepackage{hyperref}
\hypersetup{%
    unicode=true,           %
    pdftitle={Performance Analysis and Optimal Communication for ECG},    %
    pdfauthor={Lockhart, Bienz, Olson},     %
    colorlinks=true,       %
    linkcolor=green!50!black,          %
    citecolor=blue,        %
    urlcolor=blue        %
}
\usepackage[capitalise,noabbrev]{cleveref}
\crefformat{equation}{(#2#1#3)}
\usepackage{amsmath}
\numberwithin{equation}{section}
\numberwithin{figure}{section}
\numberwithin{algorithm}{section}

\usepackage{ifthen}
\usepackage{tikz}
\usetikzlibrary{arrows}
\usetikzlibrary{calc,decorations.pathreplacing,shapes.misc}
\usetikzlibrary{patterns}
\usetikzlibrary{calc,positioning}
\usepackage{import}

\usepackage{siunitx} %

\usepackage{soul} %

\newtheorem{ExampleEnv}{Example}[section]

\newcommand{\Rn}{\mathbb{R}^n}
\newcommand{\Rnn}{\mathbb{R}^{n\times n}}
\newcommand{\Rmn}{\mathbb{R}^{m\times n}}
\newcommand{\Rnt}{\mathbb{R}^{n\times t}}
\newcommand{\sproc}{s_{\text{proc}}}
\newcommand{\snode}{s_{\text{node}}}
\newcommand{\nproctonode}{m_{\text{proc$\rightarrow$node}}}
\newcommand{\nnodetonode}{m_{\text{node$\rightarrow$node}}}
\newcommand{\snodetonode}{s_{\text{node$\rightarrow$node}}}
\DeclareMathOperator{\spn}{span}

\newcommand{\ppn}{\texttt{ppn}}
\newcommand{\nnz}{\texttt{nnz}}
\newcommand{\old}{\text{\tiny old}}

\colorlet{hlgreen}{green!20}
\colorlet{hlblue}{blue!20}
\colorlet{hlyellow}{yellow!20}
\colorlet{hlred}{red!20}
\newcommand{\hlg}[1]{ \sethlcolor{hlgreen} \hl{#1} }
\newcommand{\hlb}[1]{ \sethlcolor{hlblue} \hl{#1} }
\newcommand{\hly}[1]{ \sethlcolor{hlyellow} \hl{#1} }
\newcommand{\hlr}[1]{ \sethlcolor{hlred} \hl{#1} }

\newcommand{\hlc}[2][yellow]{{%
    \colorlet{foo}{#1}%
    \sethlcolor{foo}\hl{#2}}%
}

\begin{document}

\title[Performance Analysis and Optimal Communication for ECG]{Performance Analysis and Optimal Node-Aware Communication for Enlarged Conjugate Gradient Methods}

\author{Shelby Lockhart}
\email{sll2@illinois.edu}
\orcid{0000-0003-4938-6111}
\affiliation{%
  \institution{University of Illinois at Urbana-Champaign}
  \streetaddress{Department of Computer Science}
  \city{Urbana}
  \state{Illinois}
  \postcode{61801}
  \country{USA}
}

\author{Amanda Bienz}
\email{bienz@unm.edu}
\orcid{0000-0002-8891-934X}
\affiliation{%
  \institution{University of New Mexico}
  \streetaddress{Department of Computer Science}
  \city{Albuquerque}
  \state{New Mexico}
  \postcode{87131}
  \country{USA}
}

\author{William Gropp}
\email{wgropp@illinois.edu}
\orcid{0000-0003-2905-3029}
\affiliation{%
  \institution{University of Illinois at Urbana-Champaign}
  \streetaddress{Department of Computer Science}
  \city{Urbana}
  \state{Illinois}
  \postcode{61801}
  \country{USA}
}

\author{Luke Olson}
\email{lukeo@illinois.edu}
\orcid{0000-0002-5283-6104}
\affiliation{%
  \institution{University of Illinois at Urbana-Champaign}
  \streetaddress{Department of Computer Science}
  \city{Urbana}
  \state{Illinois}
  \postcode{61801}
  \country{USA}
}

\begin{abstract}
Krylov methods are a key way of solving large sparse linear systems of equations,
but suffer from poor strong scalabilty on distributed memory
machines. This is due to high synchronization costs from large numbers of collective
communication calls alongside a low computational workload.
Enlarged Krylov methods address this issue by decreasing the
total iterations to convergence, an artifact of splitting the initial
residual and resulting in operations on block vectors.
In this paper, we present a performance study of an Enlarged
Krylov Method, Enlarged Conjugate Gradients (ECG), noting the
impact of block vectors on parallel performance at scale.
Most notably, we observe the increased overhead of point-to-point
communication as a result of denser messages in the sparse matrix-block
vector multiplication kernel. Additionally, we present models
to analyze expected performance of ECG, as well as,
motivate design decisions. Most importantly, we introduce a new point-to-point
communication approach based on node-aware communication techniques that increases
efficiency of the method at scale.
\end{abstract}



\maketitle

\section{Introduction}\label{sec:intro}

A significant performance limitation for sparse solvers on large scale
parallel computers is the lack of computational work compared to the communication
overhead~\cite{mohiyuddin2009minimizing}.
The iterative solution to large, sparse linear systems of the form $Ax=b$ often
requires \textit{many} sparse matrix-vector multiplications and
costly collective communication in the form of inner products; this is the case with
the conjugate gradient (CG) method, and with Krylov methods in general.
In this paper, we consider so-called \textit{enlarged} Krylov methods~\cite{scalable_enlarged},
wherein block vectors are introduced to improve convergence,
thereby reducing the amount of collective communication in exchange for denser point-to-point communication
in the sparse matrix-block vector multiplication.  We analyze
the associated performance expectations and introduce efficient communication methods that
render this class of methods more efficient at scale.

There have been a number of suggested algorithms for
addressing the imbalance in computation and communication
within Krylov methods, including communication
avoidance~\cite{HoemmenThesis,avoidcomm_lanzcos}, overlapping communication and
computation~\cite{pipelined}, and delaying communication at the cost of performing more
computation~\cite{s_stepIter}.
Most recently, there has been work on reducing iterations to convergence
via increasing the amount of computation per iteration, and ultimately,
the amount of data communicated~\cite{block, enlarged, sstep_enlarged}.
These approaches have been successful in reducing  the number of global
synchronization points; the current work is considered complementary in that
the goal is reduction of the total \textit{amount} of communication.

In addition to reducing synchronization points, Enlarged Krylov methods such as enlarged conjugate
gradient (ECG) reduce the number of sparse matrix-vector multiplications by improving the convergence
of the method through an increase in the amount of
computation per iteration.
This is accomplished by using block vectors, which results in both an increase in (local) computational
work, but also an increase in inter-process communication per iteration.
Consequently, the focus of this
paper is on analyzing the effects of block vectors
on the performance of ECG and proposing optimal strategies to address the communication imbalances they
introduce.

There are two key contributions made in this paper.
\begin{enumerate}
    \item A performance study and analysis of an enlarged Krylov method based on ECG,
          with an emphasis on the communication and computation
          of block vectors. Specifically, we note how they re-balance the
          point-to-point and collective communication within a single iteration
          of ECG, shifting the performance bottleneck to the point-to-point
          communication.
    \item The development of a new communication technique for blocked data based on
          node-aware communication techniques that have shown to reduce time spent in
          communication within the context of sparse matrix-vector multiplication and
          algebraic multigrid~\cite{Bienz_napspmv, Bienz_napamg}. This new
          communication technique exhibits speedups as high as 60x for various
          large-scale test matrices on two different supercomputer systems,
          as well as, reduces the point-to-point communication bottleneck in ECG\@.
\end{enumerate}
These contributions are presented in~\cref{sec:performance} and~\cref{sec:block_nap},
respectively.

\section{Background}\label{sec:background}

The conjugate gradient (CG) method for solving a system of equations, $Ax = b$,
exhibits poor parallel scalability in many
situations~\cite{ghysels2014hiding,mcinnes2014hierarchical,HoemmenThesis,avoidcomm_lanzcos}.
In particular, the \textit{strong} scalability is limited due to the high volume of
collective
communication relative to the low computational requirements of the method.
The enlarged conjugate gradient (ECG) method has a lower volume of collective communication and higher computational requirements per iteration compared to CG, thus exhibiting better strong scalability.
In this section, we detail the basic structure of ECG, briefly outlining
the method in terms of mathematical operations and highlighting the key differences
from standard CG\@. A key computation kernel in both Krylov and enlarged Krylov
methods is that of a sparse matrix-vector multiplication; we discuss node-aware
communication techniques for this operation in~\cref{sec:node_aware_comm}.

Throughout this section and the remainder of the paper, ECG performance is analyzed
with respect to the problem described in~\cref{ex:mfem}.
\begin{ExampleEnv}\label{ex:mfem}
    In this example, we consider a discontinuous Galerkin finite element discretization
    of the Laplace equation, $-\Delta u = 1$ on a unit square, with homogeneous Dirichlet boundary conditions.  The problem is generated using MFEM~\cite{mfem-library} and the
    resulting sparse matrix consists of
    $\num{1310720}$ rows and $\num{104529920}$ nonzero entries.  Graph partitioning
    is not used to reorder the entries, unless stated.
\end{ExampleEnv}

\subsection{Enlarged Krylov Subspace Methods}\label{sec:eksm}

Similar to CG, ECG
begins with an initial guess $x_0$ and seeks an update as a solution to the problem
$Ax = b$, with the initial residual given by $r_0 = b - A x_0$.
Unlike CG, which considers updates of the form $x_k \in x_0 + \mathcal{K}_{k}$, where $\mathcal{K}_{k}$ is the Krylov subspace defined as
\begin{equation}
  \mathcal{K}_{k} = \spn\{r_0, Ar_0, A^2 r_0, \dots, A^{k-1} r_0\},
\end{equation}
ECG targets $x_k \in x_0 + \mathcal{K}_{k,t}$ where $\mathcal{K}_{k,t}$ is the \textit{enlarged} Krylov space defined as
\begin{equation}\label{eq:EK}
  \mathcal{K}_{k,t} = \spn\{T_{r_0,t}, AT_{r_0,t}, A^2 T_{r_0,t}, \dots, A^{k-1} T_{r_0,t}\},
\end{equation}
with $T_{r,t}$ representing a projection of the residual $r$ (defined next).
Notably,
the enlarged Krylov subspace contains the traditional Krylov subspace: $\mathcal{K}_{k} \subset \mathcal{K}_{k,t}$~\cite{enlarged}.

\begin{figure}[!ht]
    \centering
    \small %
    \definecolor{tab-blue}{HTML}{1f77b4}
\definecolor{tab-orange}{HTML}{ff7f0e}
\definecolor{tab-green}{HTML}{2ca02c}
\definecolor{tab-red}{HTML}{d62728}
\definecolor{tab-purple}{HTML}{9467bd}
\definecolor{tab-brown}{HTML}{8c564b}
\definecolor{tab-pink}{HTML}{e377c2}
\definecolor{tab-gray}{HTML}{7f7f7f}
\definecolor{tab-olive}{HTML}{bcbd22}
\definecolor{tab-cyan}{HTML}{17becf}
\begin{tikzpicture}[x=100pt,y=100pt]
  \begin{scope}[shift={(10pt, 15pt)}]

  \foreach \y in {1,...,12}{%
    \ifnum\y<4
      \draw[fill=tab-blue,draw=tab-blue] (0, \y/10+0.01) rectangle +(0.08, 0.08);
    \else
      \ifnum\y<7
        \draw[fill=tab-red,draw=tab-red] (0, \y/10+0.01) rectangle +(0.08, 0.08);
      \else
        \ifnum\y<10
          \draw[fill=tab-green,draw=tab-green] (0, \y/10+0.01) rectangle +(0.08, 0.08);
        \else
          \draw[fill=tab-orange,draw=tab-orange] (0, \y/10+0.01) rectangle +(0.08, 0.08);
        \fi
      \fi
    \fi
  }
  \node[below, align=center] (a) at (0.05, 0.0) {$r$};
  \node[below, align=center, right= .1 of a] (b) {$\rightarrow$};
  \node[below, align=center, right= .35 of a] (c) {$T_{r,3}$};

  \foreach \z in {1, 2, 3}{%
    \foreach \y in {1,...,12}{%
      \def\x{\z/10+3/10}
      \draw[tab-gray,draw=tab-gray] (\x, \y/10+0.01) rectangle +(0.08, 0.08);

      \ifnum\y<4
        \ifnum\z=1
        \draw[fill=tab-blue,draw=tab-blue] (\x, \y/10+0.01) rectangle +(0.08, 0.08);
        \fi
      \else
        \ifnum\y<7
          \ifnum\z=3
          \draw[fill=tab-red,draw=tab-red] (\x, \y/10+0.01) rectangle +(0.08, 0.08);
          \fi
        \else
          \ifnum\y<10
            \ifnum\z=2
            \draw[fill=tab-green,draw=tab-green] (\x, \y/10+0.01) rectangle +(0.08, 0.08);
            \fi
          \else
            \ifnum\z=1
            \draw[fill=tab-orange,draw=tab-orange] (\x, \y/10+0.01) rectangle +(0.08, 0.08);
            \fi
          \fi
        \fi
      \fi
    }
  }

\begin{scope}[shift={(120pt,0)}]
  \foreach \y in {1,...,12}{%
    \ifnum\y<4
      \draw[fill=tab-blue,draw=tab-blue] (0, \y/10+0.01) rectangle +(0.08, 0.08);
    \else
      \ifnum\y<7
        \draw[fill=tab-red,draw=tab-red] (0, \y/10+0.01) rectangle +(0.08, 0.08);
      \else
        \ifnum\y<10
          \draw[fill=tab-green,draw=tab-green] (0, \y/10+0.01) rectangle +(0.08, 0.08);
        \else
          \draw[fill=tab-orange,draw=tab-orange] (0, \y/10+0.01) rectangle +(0.08, 0.08);
        \fi
      \fi
    \fi
  }
  \node[below, align=center] (a) at (0.05, 0.0) {$r$};
  \node[below, align=center, right= .1 of a] (b) {$\rightarrow$};
  \node[below, align=center, right= .35 of a] (c) {$T_{r,3}$};

  \foreach \z in {1, 2, 3}{%
    \foreach \y in {1,...,12}{%
      \def\x{\z/10+3/10}
      \draw[tab-gray,draw=tab-gray] (\x, \y/10+0.01) rectangle +(0.08, 0.08);

      \ifnum\y<4
        \ifnum\z=1
        \def\s{\x-0.1*\y-0.1}
        \draw[fill=tab-blue,draw=tab-blue] (\x+\s, \y/10+0.01) rectangle +(0.08, 0.08);
        \fi
      \else
        \ifnum\y<7
          \ifnum\z=3
          \def\s{\x-0.1*\y-0.2}
          \draw[fill=tab-red,draw=tab-red] (\x+\s, \y/10+0.01) rectangle +(0.08, 0.08);
          \fi
        \else
          \ifnum\y<10
            \ifnum\z=2
            \def\s{\x-0.1*\y+0.3}
            \draw[fill=tab-green,draw=tab-green] (\x+\s, \y/10+0.01) rectangle +(0.08, 0.08);
            \fi
          \else
            \ifnum\z=1
            \def\s{\x-0.1*\y+0.8}
            \draw[fill=tab-orange,draw=tab-orange] (\x+\s, \y/10+0.01) rectangle +(0.08, 0.08);
            \fi
          \fi
        \fi
      \fi
    }
  }
\end{scope}

\begin{scope}[shift={(240pt,0)}]
  \foreach \y in {1,...,12}{%
    \ifnum\y<4
      \draw[fill=tab-blue,draw=tab-blue] (0, \y/10+0.01) rectangle +(0.08, 0.08);
    \else
      \ifnum\y<7
        \draw[fill=tab-red,draw=tab-red] (0, \y/10+0.01) rectangle +(0.08, 0.08);
      \else
        \ifnum\y<10
          \draw[fill=tab-green,draw=tab-green] (0, \y/10+0.01) rectangle +(0.08, 0.08);
        \else
          \draw[fill=tab-orange,draw=tab-orange] (0, \y/10+0.01) rectangle +(0.08, 0.08);
        \fi
      \fi
    \fi
  }
  \node[below, align=center] (a) at (0.05, 0.0) {$r$};
  \node[below, align=center, right= .1 of a] (b) {$\rightarrow$};
  \node[below, align=center, right= .35 of a] (c) {$T_{r,3}$};

  \foreach \z in {1, 2, 3}{%
    \foreach \y in {1,...,12}{%
      \def\x{\z/10+3/10}
      \draw[tab-gray,draw=tab-gray] (\x, \y/10+0.01) rectangle +(0.08, 0.08);
    }
  }
  \draw[fill=tab-orange,draw=tab-orange] (1/10+3/10, 12/10+0.01) rectangle +(0.08, 0.08);
  \draw[fill=tab-orange,draw=tab-orange] (1/10+3/10, 11/10+0.01) rectangle +(0.08, 0.08);
  \draw[fill=tab-orange,draw=tab-orange] (2/10+3/10, 10/10+0.01) rectangle +(0.08, 0.08);
  \draw[fill=tab-green,draw=tab-green]   (2/10+3/10,  9/10+0.01) rectangle +(0.08, 0.08);
  \draw[fill=tab-green,draw=tab-green]   (3/10+3/10,  8/10+0.01) rectangle +(0.08, 0.08);
  \draw[fill=tab-green,draw=tab-green]   (3/10+3/10,  7/10+0.01) rectangle +(0.08, 0.08);
  \draw[fill=tab-red,draw=tab-red]       (1/10+3/10,  6/10+0.01) rectangle +(0.08, 0.08);
  \draw[fill=tab-red,draw=tab-red]       (1/10+3/10,  5/10+0.01) rectangle +(0.08, 0.08);
  \draw[fill=tab-red,draw=tab-red]       (2/10+3/10,  4/10+0.01) rectangle +(0.08, 0.08);
  \draw[fill=tab-blue,draw=tab-blue]     (2/10+3/10,  3/10+0.01) rectangle +(0.08, 0.08);
  \draw[fill=tab-blue,draw=tab-blue]     (3/10+3/10,  2/10+0.01) rectangle +(0.08, 0.08);
  \draw[fill=tab-blue,draw=tab-blue]     (3/10+3/10,  1/10+0.01) rectangle +(0.08, 0.08);
\end{scope}

  \end{scope}
\end{tikzpicture}

    \caption{Three examples of $T_{r,t}$, with $t=3$.  In each case, $r$ is a vector of length 12, decomposed into $12 \times 3$ block vector.  The colors represent a case of four processors: orange, green, red, and blue.}\label{fig:split_r}
\end{figure}
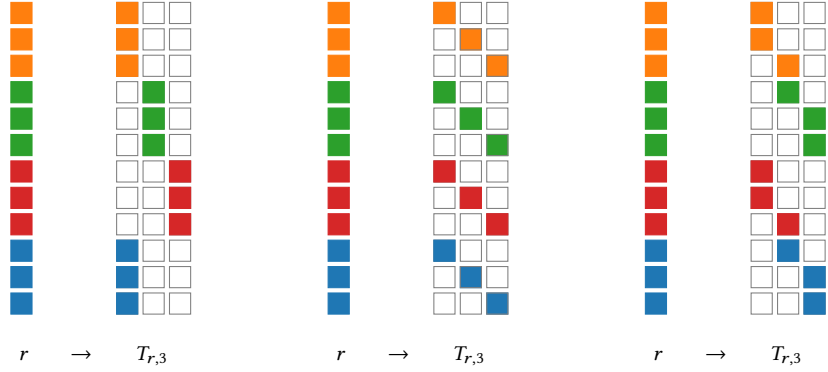

In~\cref{eq:EK}, $T_{r,t}$ defines a projection of the residual $r$ (normally the initial residual $r_0$) from $\Rn \rightarrow \Rnt$, by splitting $r$ across $t$ subdomains.
The projection may be defined in a number of ways, with the caveat that the resulting columns
of $T_{r,t}$ are linearly independent and preserve the row-sum:
\begin{equation}
    r = \sum_{i=0}^{t} (T_{r,t})_i,
\end{equation}
where we denote the $i^{\text{th}}$ column of $T$ as $(T)_i$.
An illustration of multiple permissible splittings is shown in~\cref{fig:split_r}.

Increasing the number of subdomains, $t$, increases computation from
single vector updates to block vector updates of size $n \times t$.
Additional uses of block vectors within ECG are outlined in detail in~\cref{ecg}.
On~\cref{alg:searchdir} a (small) linear system is solved to
generate the $t$ search directions.  In
addition, the sparse matrix-block vector product (SpMBV) $A P_k$ is performed at each iteration.
The number of iterations to convergence is generally reduced from that required by CG,
but the algorithm does not eliminate the communication overhead when the algorithm is
performed at scale.
Unlike CG, where the performance bottleneck is caused by the load imbalance incurred from
each inner product in the iteration, ECG sees communication overhead at scale due to the
communication associated with the SpMBV kernel (see~\cref{fig:ecg_profiling} in~\cref{sec:performance}).  We
introduce a new communication method in~\cref{sec:block_nap} to improve this performance.
\begin{algorithm2e}[!ht]
  \DontPrintSemicolon%
  $r := b-Ax$\;
  $P := 0$, $R := T_{r,t}$, $Z := R$\;
  \While{not converged}{%
      $P_{\old} := P$\;
      $P := Z (Z^T A Z)^{-\frac{1}{2}}$\;\label{alg:searchdir}
      $c := P^T R$\;\label{alg:c}
      $X := X + P c$\;
      $R := R - A P c$\;
      \If{$\| \sum_{i=1}^t (R)_i \| <$ tolerance}{%
        \texttt{break}\;
      }
      $d := (AP)^T (AP)$\;
      $d_{\old} := (AP_{\old})^T (AP)$\;
      $Z := AP - P d - P_{\old} d_{\old}$\;
  }
  $x := \sum_{i=1}^t (X)_i$\;
  \caption{Enlarged Conjugate Gradient}\label{ecg}
\end{algorithm2e}

\subsection{Node-Aware Communication Techniques}\label{sec:node_aware_comm}

Sparse matrix-vector multiplication (SpMV), defined as
\begin{equation}\label{eq:spmv}
   A \cdot v \rightarrow w
\end{equation}
with $A \in \Rmn$ and $v$, $w \in \Rn$, is a common kernel in sparse iterative methods.
It is known to lack strong scalability in a distributed memory parallel
environment, a problem stemming from low computational requirements and the communication
overhead associated with applying standard communication techniques to sparse
matrix operations.

\begin{figure}[!ht]
    \centering
    \small %
    \definecolor{tab-blue}{HTML}{1f77b4}
\definecolor{tab-orange}{HTML}{ff7f0e}
\definecolor{tab-green}{HTML}{2ca02c}
\definecolor{tab-red}{HTML}{d62728}
\definecolor{tab-purple}{HTML}{9467bd}
\definecolor{tab-brown}{HTML}{8c564b}
\definecolor{tab-pink}{HTML}{e377c2}
\definecolor{tab-gray}{HTML}{7f7f7f}
\definecolor{tab-olive}{HTML}{bcbd22}
\definecolor{tab-cyan}{HTML}{17becf}

\newcommand{\aij}[5]
{
  \draw[fill=#4!#5,draw=#4] (#2/10+3/10, 1/10+#3/10-#1/10+0.01) rectangle +(0.08, 0.08);
}

\begin{tikzpicture}[x=100pt,y=100pt]
  \begin{scope}[shift={(10pt, 15pt)}]

  \foreach \z in {1, ..., 12}{%
    \foreach \y in {1,...,12}{%
      \def\x{\z/10+3/10}
      \draw[draw=tab-gray!20] (\x, \y/10+0.01) rectangle +(0.08, 0.08);
    }
  }

  \aij{1}{1}{12}{tab-orange}{100}
  \aij{2}{2}{12}{tab-orange}{100}
  \aij{3}{3}{12}{tab-orange}{100}
  \aij{1}{3}{12}{tab-orange}{100}
  \aij{3}{2}{12}{tab-orange}{100}

  \aij{4}{4}{12}{tab-green}{100}
  \aij{5}{5}{12}{tab-green}{100}
  \aij{6}{6}{12}{tab-green}{100}
  \aij{4}{5}{12}{tab-green}{100}
  \aij{5}{4}{12}{tab-green}{100}
  \aij{5}{6}{12}{tab-green}{100}

  \aij{7}{7}{12}{tab-red}{100}
  \aij{8}{8}{12}{tab-red}{100}
  \aij{9}{9}{12}{tab-red}{100}
  \aij{9}{7}{12}{tab-red}{100}

  \aij{10}{10}{12}{tab-blue}{100}
  \aij{11}{11}{12}{tab-blue}{100}
  \aij{12}{12}{12}{tab-blue}{100}
  \aij{10}{11}{12}{tab-blue}{100}
  \aij{12}{10}{12}{tab-blue}{100}

  \aij{1}{6}{12}{tab-orange}{20}
  \aij{2}{4}{12}{tab-orange}{20}
  \aij{2}{5}{12}{tab-orange}{20}
  \aij{3}{5}{12}{tab-orange}{20}

  \aij{4}{1}{12}{tab-green}{20}
  \aij{5}{2}{12}{tab-green}{20}
  \aij{6}{1}{12}{tab-green}{20}

  \aij{7}{11}{12}{tab-red}{20}
  \aij{8}{12}{12}{tab-red}{20}
  \aij{9}{10}{12}{tab-red}{20}

  \aij{11}{9}{12}{tab-blue}{20}
  \aij{12}{8}{12}{tab-blue}{20}

  \aij{1}{10}{12}{tab-orange}{0}
  \aij{2}{8}{12}{tab-orange}{0}
  \aij{2}{9}{12}{tab-orange}{0}
  \aij{2}{12}{12}{tab-orange}{0}
  \aij{3}{7}{12}{tab-orange}{0}
  \aij{3}{10}{12}{tab-orange}{0}

  \aij{4}{9}{12}{tab-green}{0}
  \aij{5}{11}{12}{tab-green}{0}
  \aij{6}{9}{12}{tab-green}{0}

  \aij{7}{2}{12}{tab-red}{0}
  \aij{8}{4}{12}{tab-red}{0}
  \aij{9}{3}{12}{tab-red}{0}

  \aij{10}{5}{12}{tab-blue}{0}
  \aij{11}{4}{12}{tab-blue}{0}
  \aij{12}{1}{12}{tab-blue}{0}

  \begin{scope}[shift={(1.4,0)}]
  \aij{1}{1}{12}{tab-orange}{100}
  \aij{2}{1}{12}{tab-orange}{100}
  \aij{3}{1}{12}{tab-orange}{100}
  \aij{4}{1}{12}{tab-green}{100}
  \aij{5}{1}{12}{tab-green}{100}
  \aij{6}{1}{12}{tab-green}{100}
  \aij{7}{1}{12}{tab-red}{100}
  \aij{8}{1}{12}{tab-red}{100}
  \aij{9}{1}{12}{tab-red}{100}
  \aij{10}{1}{12}{tab-blue}{100}
  \aij{11}{1}{12}{tab-blue}{100}
  \aij{12}{1}{12}{tab-blue}{100}
  \end{scope}

  \begin{scope}[shift={(1.8,0)}]
  \aij{1}{1}{12}{tab-orange}{100}
  \aij{2}{1}{12}{tab-orange}{100}
  \aij{3}{1}{12}{tab-orange}{100}
  \aij{4}{1}{12}{tab-green}{100}
  \aij{5}{1}{12}{tab-green}{100}
  \aij{6}{1}{12}{tab-green}{100}
  \aij{7}{1}{12}{tab-red}{100}
  \aij{8}{1}{12}{tab-red}{100}
  \aij{9}{1}{12}{tab-red}{100}
  \aij{10}{1}{12}{tab-blue}{100}
  \aij{11}{1}{12}{tab-blue}{100}
  \aij{12}{1}{12}{tab-blue}{100}
  \end{scope}

  \node[below, align=center] (a) at (1.0, 0.05) {$A$};
  \node[below, align=center, right= 0.55 of a] (b) {$*$};
  \node[below, align=center, right= .04 of b] (c) {$v$};
  \node[below, align=center, right= .04 of c] (d) {$\rightarrow$};
  \node[below, align=center, right= .04 of d] (e) {$w$};

  \draw[black, solid] (0, 0.70) -- (2.5, .70);
  \node[rotate=90,anchor=north] at (0.0, 1.05) {node $0$};
  \node[rotate=90,anchor=north] at (0.0, .35) {node $1$};

  \draw[black, dashed] (0.2, 1.0) -- (2.4, 1.0);
  \draw[black, dashed] (0.2, 0.4) -- (2.4, 0.4);
  \node[anchor=north] at (0.25, 1.25) {p0};
  \node[anchor=north] at (0.25, 0.95) {p1};
  \node[anchor=north] at (0.25, 0.65) {p2};
  \node[anchor=north] at (0.25, 0.35) {p3};

  \end{scope}
\end{tikzpicture}

    \caption{Communication and partitioning of a SpMV, $A \cdot v\rightarrow w$.
      With $n=12$, matrix $A$
      and vectors $v$ and $w$ are partitioned across two nodes and four processors (p0, p1, p2, and p3),
    indicated by the colors orange, green, red, and blue.
    A solid block,~\protect\tikz{\protect\draw[fill=black] (0,0) rectangle (0.2,0.2);},
    represents the portion of the SpMV requiring only on-process values
    from $v$. A shaded block,~\protect\tikz{\protect\draw[fill=black!20] (0,0) rectangle (0.2,0.2);},
    represents the portion of the SpMV requiring
    on-node but off-process communication of values from $v$, and the outlined
    blocks,~\protect\tikz{\protect\draw[fill=black!0] (0,0) rectangle (0.2,0.2);},
    require values of $v$ from processors off-node.}\label{fig:spmv_nodeaware}
\end{figure}
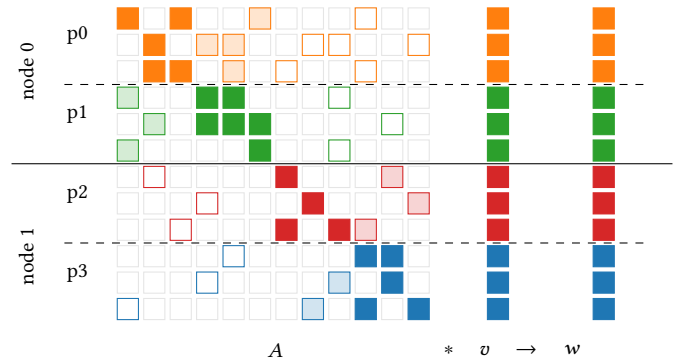

Generally, $A$, $v$, and $w$ are partitioned row-wise across
$p$ processes with contiguous rows stored on each process (see~\cref{fig:spmv_nodeaware}).
In addition, we split the rows of $A$ on each process into 2 blocks, namely
on-process and off-process.
The on-process block is the diagonal block of columns corresponding to the
on-process portion of rows in $v$ and $w$, and the off-process block contains
the matrix $A$'s nonzero values correlating to non-local rows of $v$ and $w$ stored
off-process. This splitting is common practice, as it differentiates between the
portions of a SpMV that require communication with other processes.

\begin{figure}[!ht]
\centering
\includegraphics{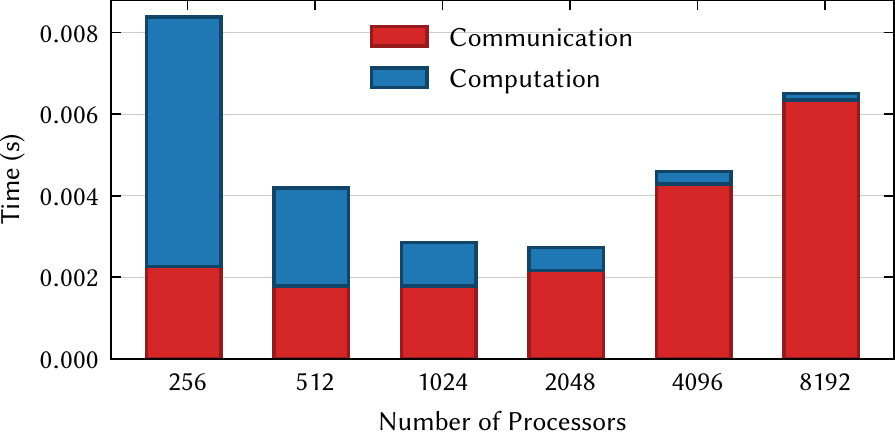}
\caption{The time required for a single
SpMV, split into communication and computation, for~\cref{ex:mfem} run on
Blue Waters with 16 processes per node.}\label{fig:spmv_breakdown}
\end{figure}

A common approach to a SpMV is to compute the local portion of the SpMV with
the on-process block of $A$ while messages are exchanged to attain the off-process
portions of $v$ necessary for the local update of $w$. While this allows for the
overlap of some communication and computation, it requires the exchange of many
point-to-point messages, which still creates a large communication overhead (see~\cref{fig:spmv_breakdown}).

\begin{figure}[!ht]
    \small %
\definecolor{tab-blue}{HTML}{1f77b4}
\definecolor{tab-orange}{HTML}{ff7f0e}
\definecolor{tab-green}{HTML}{2ca02c}
\definecolor{tab-red}{HTML}{d62728}
\definecolor{tab-purple}{HTML}{9467bd}
\definecolor{tab-brown}{HTML}{8c564b}
\definecolor{tab-pink}{HTML}{e377c2}
\definecolor{tab-gray}{HTML}{7f7f7f}
\definecolor{tab-olive}{HTML}{bcbd22}
\definecolor{tab-cyan}{HTML}{17becf}

\newcommand{\anode}
{
  \begin{scope}[x=40pt,y=40pt]
  \draw[draw=tab-gray!80!black, fill=tab-gray!30] (0,     0) rectangle +(2, 2);
  \draw[draw=tab-orange!80!black,fill=tab-orange!25]  (0.5, 1.5) circle (0.35 and .45);
  \draw[draw=tab-green!80!black, fill=tab-green!25]   (1.5, 1.5) circle (0.35 and .45);
  \draw[draw=tab-red!80!black,   fill=tab-red!25]     (0.5, 0.5) circle (0.35 and .45);
  \draw[draw=tab-blue!80!black,  fill=tab-blue!25]    (1.5, 0.5) circle (0.35 and .45);
  \coordinate (p0) at (0.5, 1.4);
  \coordinate (p1) at (1.5, 1.4);
  \coordinate (p2) at (0.5, 0.4);
  \coordinate (p3) at (1.5, 0.4);
  \node[anchor=center] at (p0) {\scriptsize P0};
  \node[anchor=center] at (p1) {\scriptsize P1};
  \node[anchor=center] at (p2) {\scriptsize P2};
  \node[anchor=center] at (p3) {\scriptsize P3};
  \node[anchor=south] at (1.0, 2.0) {Node 0};

  \begin{scope}[shift={(90pt,0pt)}]
  \draw[dashed,draw=tab-gray!80!black, fill=tab-gray!30] (0,     0) rectangle +(2, 2);
  \draw[dashed,draw=tab-orange!80!black,fill=tab-orange!25]  (0.5, 1.5) circle (0.35 and .45);
  \draw[dashed,draw=tab-green!80!black, fill=tab-green!25]   (1.5, 1.5) circle (0.35 and .45);
  \draw[dashed,draw=tab-red!80!black,   fill=tab-red!25]     (0.5, 0.5) circle (0.35 and .45);
  \draw[dashed,draw=tab-blue!80!black,  fill=tab-blue!25]    (1.5, 0.5) circle (0.35 and .45);
  \coordinate (p4) at (0.5, 1.4);
  \coordinate (p5) at (1.5, 1.4);
  \coordinate (p6) at (0.5, 0.4);
  \coordinate (p7) at (1.5, 0.4);
  \node[anchor=center] at (p4) {\scriptsize P4};
  \node[anchor=center] at (p5) {\scriptsize P5};
  \node[anchor=center] at (p6) {\scriptsize P6};
  \node[anchor=center] at (p7) {\scriptsize P7};
  \node[anchor=south] at (1.0, 2.0) {Node 1};
  \end{scope}
  \end{scope}
}

\begin{tikzpicture}[x=80pt,y=80pt]
  \begin{scope}[shift={(3pt, 15pt)}]
  \anode
  \tikzstyle{myarrow}=[>=latex,shorten >=5pt,shorten <=5pt,->,draw,thin,color=black]

  \path[myarrow] (p0) to (p6);
  \path[myarrow] (p1) to (p6);
  \path[myarrow] ([yshift=-10pt]p2.south) to (p6);
  \path[myarrow] (p3) to (p6);
  \path[myarrow] (p4) to (p6);
  \path[myarrow] (p5) to (p6);
  \path[myarrow] (p7) to (p6);

  \node at (85pt, -12.0pt) {Sending multiple messages to Node 1};

  \end{scope}
  \begin{scope}[shift={(200pt, 15pt)}]
  \anode
  \tikzstyle{myarrow}=[>=latex,shorten >=5pt,shorten <=0pt,->,draw,thin,color=black]

  \foreach \j in {-4,0,4}{
    \draw[color=black!70, fill=tab-gray] ([xshift=\j pt,yshift=5pt]p1) rectangle +(0.04, 0.04);
  }
  \coordinate (myp) at ([xshift=9pt,yshift=6pt] p1);
  \path[myarrow] (myp) to[bend left] (p4);
  \path[myarrow] (myp) to[bend left] (p5);
  \path[myarrow] (myp) to[looseness=.7, bend right=45] (p6);
  \path[myarrow] (myp) to[bend right=75,in=90,out=-45] (p7);
  \node at (85pt, -12.0pt) {Sending duplicate data to Node 1};

  \end{scope}
\end{tikzpicture}

    \caption{Standard communication.
    On the left, Node 0
    injects multiple messages into the network, all to P6 on Node 1.
    On the right, P1 sends multiple messages containing the same data to each process
  on Node 1, resulting in "duplicate" data being sent across the network.}\label{fig:standard_comm}
\end{figure}
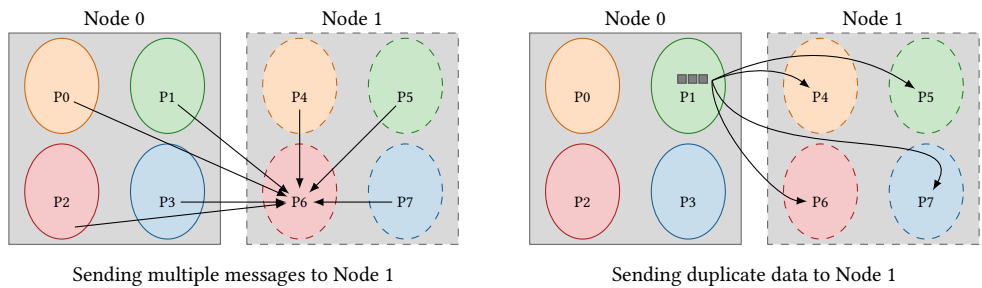

The inefficiency of this standard approach is attributed to two
redundancies shown in~\cref{fig:standard_comm}.
First, many messages are injected into the network from each node.
Some nodes are sending multiple messages to a single destination process on a
separate node, creating a redundancy of messages.
Secondly, processes send the necessary values from their local portion of $v$
to any other process requiring
that information for its local computation of $w$. However, the same information may already be
available and received by a separate process
on the same node, creating a redundancy of data being sent through the network.

Node-aware communication techniques~\cite{Bienz_napspmv, Bienz_napamg}
mitigate these issues by considering node
topology to further break down the off-process block into vector values of $v$
that require on- or off-node communication; this decomposition
is shown in~\cref{fig:spmv_nodeaware}.
As a result,
costly redundant messages are traded for faster, on-node communication,
resulting in 2 different multi-step schemes, namely 3-step and 2-step.

\subsubsection*{3-step Node-Aware Communication}

3-step node-aware communication eliminates both redundancies in
standard communication by gathering all necessary data to be sent off-node in
a single buffer. Efficient implementation of this method relies
on pairing all processes with a receiving process on distinct nodes -- ensuring
that every process remains active throughout the entire communication scheme.

\begin{figure}[!ht]
    \centering
    \small %
    \definecolor{tab-blue}{HTML}{1f77b4}
\definecolor{tab-orange}{HTML}{ff7f0e}
\definecolor{tab-green}{HTML}{2ca02c}
\definecolor{tab-red}{HTML}{d62728}
\definecolor{tab-purple}{HTML}{9467bd}
\definecolor{tab-brown}{HTML}{8c564b}
\definecolor{tab-pink}{HTML}{e377c2}
\definecolor{tab-gray}{HTML}{7f7f7f}
\definecolor{tab-olive}{HTML}{bcbd22}
\definecolor{tab-cyan}{HTML}{17becf}

\newcommand{\anode}
{
  \begin{scope}[x=40pt,y=40pt]
  \draw[draw=tab-gray!80!black, fill=tab-gray!30] (0,     0) rectangle +(2, 2);
  \draw[draw=tab-orange!80!black,fill=tab-orange!25]  (0.5, 1.5) circle (0.35 and .45);
  \draw[draw=tab-green!80!black, fill=tab-green!25]   (1.5, 1.5) circle (0.35 and .45);
  \draw[draw=tab-red!80!black,   fill=tab-red!25]     (0.5, 0.5) circle (0.35 and .45);
  \draw[draw=tab-blue!80!black,  fill=tab-blue!25]    (1.5, 0.5) circle (0.35 and .45);
  \coordinate (p0) at (0.5, 1.4);
  \coordinate (p1) at (1.5, 1.4);
  \coordinate (p2) at (0.5, 0.4);
  \coordinate (p3) at (1.5, 0.4);
  \node[anchor=center] at (p0) {\scriptsize P0};
  \node[anchor=center] at (p1) {\scriptsize P1};
  \node[anchor=center] at (p2) {\scriptsize P2};
  \node[anchor=center] at (p3) {\scriptsize P3};
  \node[anchor=south] at (1.0, 2.0) {Node 0};

  \begin{scope}[shift={(90pt,0pt)}]
  \draw[dashed,draw=tab-gray!80!black, fill=tab-gray!30] (0,     0) rectangle +(2, 2);
  \draw[dashed,draw=tab-orange!80!black,fill=tab-orange!25]  (0.5, 1.5) circle (0.35 and .45);
  \draw[dashed,draw=tab-green!80!black, fill=tab-green!25]   (1.5, 1.5) circle (0.35 and .45);
  \draw[dashed,draw=tab-red!80!black,   fill=tab-red!25]     (0.5, 0.5) circle (0.35 and .45);
  \draw[dashed,draw=tab-blue!80!black,  fill=tab-blue!25]    (1.5, 0.5) circle (0.35 and .45);
  \coordinate (p4) at (0.5, 1.4);
  \coordinate (p5) at (1.5, 1.4);
  \coordinate (p6) at (0.5, 0.4);
  \coordinate (p7) at (1.5, 0.4);
  \node[anchor=center] at (p4) {\scriptsize P4};
  \node[anchor=center] at (p5) {\scriptsize P5};
  \node[anchor=center] at (p6) {\scriptsize P6};
  \node[anchor=center] at (p7) {\scriptsize P7};
  \node[anchor=south] at (1.0, 2.0) {Node 1};
  \end{scope}
  \end{scope}
}

\tikzset{
  doublearrowp/.style args={#1 and #2}{
    round cap-latex,line width=1pt,#1,shorten <=5pt,
    postaction={draw,round cap-latex,#2,line width=1pt/3,shorten <=5.5pt,shorten >=1.5pt},
  },
  doublearrowq/.style args={#1 and #2}{
    round cap-latex,line width=1pt,#1, shorten <=2pt, shorten >=5pt,
    postaction={draw,round cap-latex,#2,line width=1pt/3,shorten <=2.5pt,shorten >=6.5pt},
  },
  doublearrowz/.style args={#1 and #2}{
    round cap-latex,line width=1pt,#1,shorten <=2pt, shorten >=2pt,
    postaction={draw,round cap-latex,#2,line width=1pt/3,shorten <=2.5pt,shorten >=3.5pt},
  },
  doublearrowp2/.style args={#1}{
    round cap-latex,line width=1pt,#1,shorten <=5pt,
  },
  doublearrowq2/.style args={#1}{
    round cap-latex,line width=1pt,#1, shorten <=2pt, shorten >=5pt,
  },
  doublearrowz2/.style args={#1}{
    round cap-latex,line width=1pt,#1,shorten <=2pt, shorten >=5pt,
  },
}
\begin{tikzpicture}[x=80pt,y=80pt]
  \begin{scope}[shift={(3pt, 5pt)}]
  \newcommand{\somenode}[3]{
  \begin{scope}
  \anode

  \foreach \j in {-4,0,4}{
    \draw[color=black!70, fill=tab-gray] ([xshift=\j pt,yshift=5pt]p0) rectangle +(0.04, 0.04);
  }
  \coordinate (myp) at ([xshift=9pt,yshift=6pt] p0);
  \draw[doublearrowp2=#1] (p1) to (myp);
  \draw[doublearrowp2=#1] (p2) to (myp);
  \draw[doublearrowp2=#1] (p3) to (myp);

  \foreach \j in {-4,0,4}{
    \draw[color=black!70, fill=tab-gray] ([xshift=\j pt,yshift=5pt]p7) rectangle +(0.04, 0.04);
  }
  \coordinate (myq) at ([xshift=-4pt,yshift=6pt] p7);
  \draw[doublearrowq2=#2] (myq) to (p4);
  \draw[doublearrowq2=#2] (myq) to (p5);
  \draw[doublearrowq2=#2] (myq) to (p6);

  \foreach \j in {-4,0,4}{
    \draw[color=black!70, fill=tab-gray] ([xshift=\j pt,yshift=5pt]p7) rectangle +(0.04, 0.04);
  }

  \draw[doublearrowz2=#3] (myp) to[out=80, in=160] (myq);
    \end{scope}
  }

  \begin{scope}[xshift=20pt,yshift=200pt]
    \node[rotate=90] at (-10pt,40pt) {Step 1};
    \somenode{dotted}{solid}{solid}
  \end{scope}
  \begin{scope}[xshift=100pt,yshift=100pt]
    \node[rotate=90] at (-10pt,40pt) {Step 2};
    \somenode{solid}{solid}{dotted}
  \end{scope}
  \begin{scope}[xshift=180pt,yshift=0pt]
    \node[rotate=90] at (-10pt,40pt) {Step 3};
    \somenode{solid}{dotted}{solid}
  \end{scope}

  \end{scope}
\end{tikzpicture}

    \caption{3-step node-aware.
    In Step 1, all the data on
    Node 0 that needs to be sent to Node 1 is collected in a buffer
    on P0, the process paired to send and receive from Node 1.
    In Step 2, P0 sends this buffer from Node 0 to P7,
    to the receiving process on Node 1. In Step 3, P7
    redistributes the data to the correct receiving processes on Node 1.}\label{fig:3step}
\end{figure}
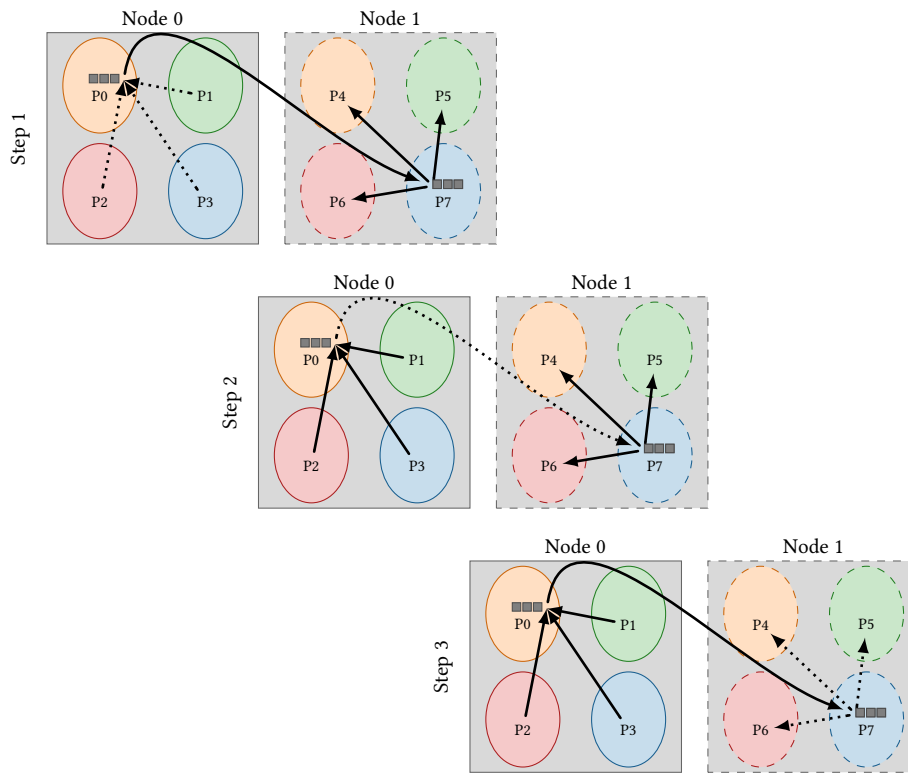

First, all data to be sent to a separate node are
gathered in a buffer by the single process paired with that node.
Secondly, this process sends the data buffer to the paired process on the
receiving node. Thirdly, the paired process on the receiving node redistributes
the data locally to the correct destination processes on-node. An example of these steps
is outlined in~\cref{fig:3step}.

\subsubsection*{2-Step Node-Aware Communication}

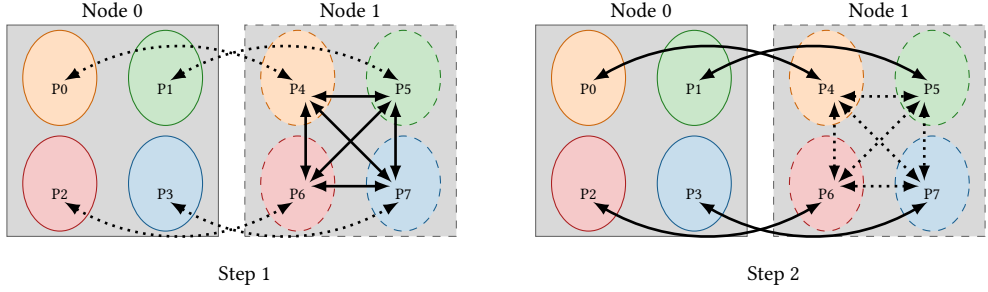
\begin{figure}[!htb]
    \centering
    \small %
    \definecolor{tab-blue}{HTML}{1f77b4}
\definecolor{tab-orange}{HTML}{ff7f0e}
\definecolor{tab-green}{HTML}{2ca02c}
\definecolor{tab-red}{HTML}{d62728}
\definecolor{tab-purple}{HTML}{9467bd}
\definecolor{tab-brown}{HTML}{8c564b}
\definecolor{tab-pink}{HTML}{e377c2}
\definecolor{tab-gray}{HTML}{7f7f7f}
\definecolor{tab-olive}{HTML}{bcbd22}
\definecolor{tab-cyan}{HTML}{17becf}

\newcommand{\anode}
{
  \begin{scope}[x=40pt,y=40pt]
  \draw[draw=tab-gray!80!black, fill=tab-gray!30] (0,     0) rectangle +(2, 2);
  \draw[draw=tab-orange!80!black,fill=tab-orange!25]  (0.5, 1.5) circle (0.35 and .45);
  \draw[draw=tab-green!80!black, fill=tab-green!25]   (1.5, 1.5) circle (0.35 and .45);
  \draw[draw=tab-red!80!black,   fill=tab-red!25]     (0.5, 0.5) circle (0.35 and .45);
  \draw[draw=tab-blue!80!black,  fill=tab-blue!25]    (1.5, 0.5) circle (0.35 and .45);
  \coordinate (p0) at (0.5, 1.4);
  \coordinate (p1) at (1.5, 1.4);
  \coordinate (p2) at (0.5, 0.4);
  \coordinate (p3) at (1.5, 0.4);
  \node[anchor=center] at (p0) {\scriptsize P0};
  \node[anchor=center] at (p1) {\scriptsize P1};
  \node[anchor=center] at (p2) {\scriptsize P2};
  \node[anchor=center] at (p3) {\scriptsize P3};
  \node[anchor=south] at (1.0, 2.0) {Node 0};

  \begin{scope}[shift={(90pt,0pt)}]
  \draw[dashed,draw=tab-gray!80!black, fill=tab-gray!30] (0,     0) rectangle +(2, 2);
  \draw[dashed,draw=tab-orange!80!black,fill=tab-orange!25]  (0.5, 1.5) circle (0.35 and .45);
  \draw[dashed,draw=tab-green!80!black, fill=tab-green!25]   (1.5, 1.5) circle (0.35 and .45);
  \draw[dashed,draw=tab-red!80!black,   fill=tab-red!25]     (0.5, 0.5) circle (0.35 and .45);
  \draw[dashed,draw=tab-blue!80!black,  fill=tab-blue!25]    (1.5, 0.5) circle (0.35 and .45);
  \coordinate (p4) at (0.5, 1.4);
  \coordinate (p5) at (1.5, 1.4);
  \coordinate (p6) at (0.5, 0.4);
  \coordinate (p7) at (1.5, 0.4);
  \node[anchor=center] at (p4) {\scriptsize P4};
  \node[anchor=center] at (p5) {\scriptsize P5};
  \node[anchor=center] at (p6) {\scriptsize P6};
  \node[anchor=center] at (p7) {\scriptsize P7};
  \node[anchor=south] at (1.0, 2.0) {Node 1};
  \end{scope}
  \end{scope}
}

\tikzset{
  doublearrowp/.style args={#1 and #2}{
    latex-latex,line width=1pt,#1,                   shorten <=2pt,  shorten >=2pt,
    postaction={draw,latex-latex,#2,line width=1pt/3,shorten <=3.5pt,shorten >=3.5pt},
  },
  doublearrowq/.style args={#1 and #2}{
    latex-latex,line width=1pt,#1,                   shorten <=2pt,  shorten >=2pt,
    postaction={draw,latex-latex,#2,line width=1pt/3,shorten <=3.5pt,shorten >=3.5pt},
  },
  doublearrowp2/.style args={#1}{
    latex-latex,line width=1pt,#1,                   shorten <=2pt,  shorten >=2pt,
  },
  doublearrowq2/.style args={#1}{
    latex-latex,line width=1pt,#1,                   shorten <=2pt,  shorten >=2pt,
  },
}
\begin{tikzpicture}[x=80pt,y=80pt]
  \begin{scope}[shift={(3pt, 25pt)}]
  \newcommand{\somenode}[2]{
  \begin{scope}
  \anode

  \draw[doublearrowp2=#1] ([xshift= 3pt, yshift=-3pt]p4) to ([xshift=-3pt, yshift=-3pt]p5);
  \draw[doublearrowp2=#1] ([xshift=-3pt, yshift=-3pt]p5) to ([xshift=-3pt, yshift= 3pt]p7);
  \draw[doublearrowp2=#1] ([xshift=-3pt, yshift= 3pt]p7) to ([xshift= 3pt, yshift= 3pt]p6);
  \draw[doublearrowp2=#1] ([xshift= 3pt, yshift= 3pt]p6) to ([xshift= 3pt, yshift=-3pt]p4);
  \draw[doublearrowp2=#1] ([xshift= 3pt, yshift=-3pt]p4) to ([xshift=-3pt, yshift= 3pt]p7);
  \draw[doublearrowp2=#1] ([xshift=-3pt, yshift=-3pt]p5) to ([xshift= 3pt, yshift= 3pt]p6);

  \draw[doublearrowq2=#2, bend left]  ([yshift=2pt]p0) to ([yshift=2pt]p4);
  \draw[doublearrowq2=#2, bend left]  ([yshift=2pt]p1) to ([yshift=2pt]p5);
  \draw[doublearrowq2=#2, bend right] ([yshift=-2pt]p2) to ([yshift=-2pt]p6);
  \draw[doublearrowq2=#2, bend right] ([yshift=-2pt]p3) to ([yshift=-2pt]p7);

  \end{scope}
  }

  \begin{scope}[xshift=0pt,yshift=0pt]
    \node at (90pt,-15pt) {Step 1};
    \somenode{solid}{dotted}
  \end{scope}
  \begin{scope}[xshift=200pt,yshift=0pt]
    \node at (90pt,-15pt) {Step 2};
    \somenode{dotted}{solid}
  \end{scope}

  \end{scope}
\end{tikzpicture}

    \caption{2-Step node-aware.
    Each process on Node 0
    is paired with a receiving process on Node 1. In Step 1, each
    process on Node 0 sends the data needed by any process on Node 1
    to its paired process on Node 1. Here, P0 is sending to P4, P1
    to P5, P2 to P6, and P3 to P7. In Step 2, each process on Node 1
    redistributes the data received from Node 0 to the destination on
    on Node 1.}\label{fig:2step}
\end{figure}

2-step node-aware communication eliminates the redundancy of sending
duplicate data from standard communication and decreases the number of inter-node
messages, but not to the same degree as 3-step communication.
In this case, \textit{each}
process exchanges the information
needed by the receiving node with their paired process directly.
Then the receiving node redistributes the
messages on-node as shown in~\cref{fig:2step}.
While multiple messages are sent to the
same node, the duplicate data being sent through the network is
eliminated. Hence, the number of bytes communicated with 3-step and 2-step
node-aware schemes is the same, and often yields a significant reduction over
the amount of data being sent through the network with standard
communication.

\subsubsection{Node-Aware Communication Models}\label{sec:node_aware_models}

\begin{table}
  \begin{tabular}{lll}
    \toprule
    parameter & description & first use\\
    \midrule
    $p$            & number of processes                                & \S\ref{sec:node_aware_comm}\\
    \nnz\          & number of nonzeros in $A$                          & \S\ref{sec:performance}\\
    $\alpha$       & network latency                                    & (\ref{eq:max_rate})\\
    $s$            & maximum number of bytes sent by a process          & (\ref{eq:max_rate})\\
    $m$            & maximum number of messages sent by a process       & (\ref{eq:max_rate})\\
    \ppn\          & processes per node                                 & (\ref{eq:max_rate})\\
    $R_N$          & network injection rate (B/s)                       & (\ref{eq:max_rate})\\
    $R_b$          & network rate (B/s)                                 & (\ref{eq:max_rate})\\
    \midrule
    $\sproc$       & maximum number of bytes sent by a process          & (\ref{eq:model2})\\
    $\snode$       & maximum number of bytes injected by a node         & (\ref{eq:model2})\\
    $\nproctonode$ & maximum number of nodes to which a processor sends & (\ref{eq:model2})\\
    $\nnodetonode$ & maximum number of messages between two nodes       & (\ref{eq:model3})\\
    $\snodetonode$ & maximum size of a message between two nodes        & (\ref{eq:model3})\\
    \bottomrule
  \end{tabular}
  \caption{Modeling parameters.}\label{tab:parameters}
\end{table}

The \textit{max-rate} model~\cite{maxrate_model} is used
to quantify the efficiency of node-aware communication
throughout the remainder of~\cref{sec:background,sec:performance}.
For clarity, all modeling
parameters referenced throughout the remainder of the paper
are defined in~\cref{tab:parameters}.
The max-rate model is an improvement to the standard postal model of communication,
accounting for injection limits into the network.
The cost of sending messages
from a symmetric multiprocessing (SMP) node is modeled as
\begin{equation}\label{eq:max_rate}
    T = \alpha \cdot m +
    \max\left(
              \frac{\ppn \cdot s}{R_N}, \frac{s}{R_b}
        \right)
\end{equation}
where $\alpha$ is the latency, $m$ is the number of messages sent by a
single process on a given node,
$s$ is the number of bytes sent by a single process on a given SMP node,
\ppn\ is the number of processes per node,
$R_N$ is the rate at which a network interface card (NIC) can inject data
into the network, and $R_b$ is the rate at which a process can transport data.

In the case of on-node messages, the injection rate is not present and
the max-rate model reduces
to the standard postal model for communication
\begin{equation}\label{eq:postal_model}
  T = \alpha_{\ell} \cdot m + \frac{s}{R_{b,\ell}},
\end{equation}
where $\alpha_{\ell}$ is the \textit{local} or on-node latency and
$R_{b,\ell}$ is the rate of sending a message on-node.

In~\cite{Bienz_napamg}, the max-rate model is extended to 2-step and 3-step
communication by splitting the model into inter-node and intra-node
components.
For 3-step, the communication model becomes
\begin{equation}\label{eq:model3}
  T_{\text{total}} =
  \underbrace{%
              \alpha \cdot \frac{\nnodetonode}{\ppn} +
              \max\left(\frac{\snode}{R_N}, \frac{\sproc}{R_b}\right)
             }_{\text{inter-node}}
+ \underbrace{%
              2\cdot\left(
                  \alpha_{\ell} \cdot (\ppn - 1) +
                  \frac{\snodetonode}{R_{b,\ell}}
                    \right)
              }_{\text{intra-node}}
\end{equation}
where $\nnodetonode$ is the maximum number of messages
communicated between any two nodes and $\snodetonode$ is
the size of messages communicated between any two nodes.

For 2-step, this results in
\begin{equation}\label{eq:model2}
  T_{\text{total}} =
  \underbrace{%
              \alpha \cdot \nproctonode +
              \max\left(\frac{\snode}{R_N}, \frac{\sproc}{R_b}\right)
             }_{\text{inter-node}}
+ \underbrace{%
              \alpha_{\ell} \cdot (\ppn - 1) +
              \frac{\sproc}{R_{b,\ell}}
             }_{\text{intra-node}}
\end{equation}
where $\snode$ and $\sproc$ represent the maximum
number of bytes injected by a single NIC and communicated by a single
process from an SMP node, respectively, and $\nproctonode$ is the maximum
number of nodes with which any process communicates.

The latency to communicate between nodes, $\alpha$, is often much higher
than the intra-node latency, $\alpha_{\ell}$, thus motivating
a multi-step communication approach.
In a 2-step method, having every process on-node communicate data
minimizes the constant factor
$\max\left(\frac{\snode}{R_N}, \frac{\sproc}{R_b}\right)$,
which depends on the maximum amount of data being communicated to
a separate node by a single process.
In practice, a 3-step method often yields the best performance for a parallel SpMV
since
the amount of data being communicated
by a single process is often small.  As a result moving the data to be
communicated off-node into a single buffer minimizes the first term
in~\cref{eq:model3}.
These multi-step communication techniques minimize the amount of time spent in
inter-node communication.  We extend this idea to the block vector operation
in~\cref{sec:blockmodel}.

\section{Performance Study of Enlarged Conjugate Gradient}\label{sec:performance}

In this section we detail the
per-iteration
performance and performance modeling of ECG\@.
A communication efficient version of~\cref{ecg} is implemented in Raptor~\cite{raptor} and is based on the work in~\cite{scalable_enlarged}.
Throughout this section and the remainder of the paper, we assume an
$n\times n$ matrix $A$ with \nnz\ nonzeros is
partitioned row-wise across a set of $p$ processes.
Each process contains at most
$\frac{n}{p}$ contiguous rows of the matrix $A$.
In the modeling that follows, we assume
a equal number of nonzeros per partition.
In addition, each block vector in~\cref{ecg}~---~$R$, $X$, $Z$, $AZ$, $P$, and $AP$~---~is partitioned
row-wise, and with the same row distribution as $A$.
The variables $c$, $d$, and
$d_{\old}$ are size $t \times t$ and a copy of each is stored locally on
each process. Tests were performed on the Blue Waters Cray XE/XK machine at the
National Center for Supercomputing Applications (NCSA) at University of Illinois.
Blue Waters~\cite{BW1,BW2} contains a 3D torus Gemini interconnect; each Gemini consists of
two nodes. The complete system contains 22 636 XE compute nodes; each
with two AMD 6276 Interlagos processors, and additional XK compute nodes unused
in these tests.

\subsection{Implementation}

\begin{figure}[!ht]
    \centering
    \small %
    \definecolor{tab-blue}{HTML}{1f77b4}
\definecolor{tab-orange}{HTML}{ff7f0e}
\definecolor{tab-green}{HTML}{2ca02c}
\definecolor{tab-red}{HTML}{d62728}
\definecolor{tab-purple}{HTML}{9467bd}
\definecolor{tab-brown}{HTML}{8c564b}
\definecolor{tab-pink}{HTML}{e377c2}
\definecolor{tab-gray}{HTML}{7f7f7f}
\definecolor{tab-olive}{HTML}{bcbd22}
\definecolor{tab-cyan}{HTML}{17becf}

\newcommand{\blockvector}[1]{
  \begin{scope}[shift={(#1pt,0)}]
  \foreach \y in {1,...,12}{%
    \ifnum\y<4
      \draw[fill=tab-blue,draw=tab-blue] (0, \y/10+0.01) rectangle +(0.08, 0.08);
    \else
      \ifnum\y<7
        \draw[fill=tab-red,draw=tab-red] (0, \y/10+0.01) rectangle +(0.08, 0.08);
      \else
        \ifnum\y<10
          \draw[fill=tab-green,draw=tab-green] (0, \y/10+0.01) rectangle +(0.08, 0.08);
        \else
          \draw[fill=tab-orange,draw=tab-orange] (0, \y/10+0.01) rectangle +(0.08, 0.08);
        \fi
      \fi
    \fi
  }
  \end{scope}
}
\newcommand{\blockblock}[3]{
  \begin{scope}[shift={(#1pt,#2pt)}]
  \foreach \x in {0,...,1}{%
    \foreach \y in {0,...,1}{%
      \ifthenelse{\equal{#3}{solid}}
        {
          \draw[fill=tab-gray,draw=black] (\x/10, \y/10) rectangle +(0.08, 0.08);
        }
        {
        \draw[pattern=#3,pattern color=tab-gray,draw=black] (\x/10, \y/10) rectangle +(0.08, 0.08);
        }
    }
  }
  \end{scope}
}
\begin{tikzpicture}[x=100pt,y=100pt]
  \begin{scope}[shift={(10pt, 15pt)}]
  \draw[black, solid]   (0, 0.70) -- (3.8, .70);
  \draw[black, dashed] (0.2, 1.0) -- (3.7, 1.0);
  \draw[black, dashed] (0.2, 0.4) -- (3.7, 0.4);
  \node[rotate=90,anchor=north] at (0.0, 1.05) {node $0$};
  \node[rotate=90,anchor=north] at (0.0, .35) {node $1$};
  \node[anchor=north] at (0.25, 1.25) {p0};
  \node[anchor=north] at (0.25, 0.95) {p1};
  \node[anchor=north] at (0.25, 0.65) {p2};
  \node[anchor=north] at (0.25, 0.35) {p3};

  \foreach \y in {105,15,45,75}{%
    \blockblock{290}{\y}{solid}
    \blockblock{320}{\y}{crosshatch dots}
    \blockblock{350}{\y}{crosshatch}
  }
  \node[anchor=south] at (3.0, -0.05) {$c$};
  \node[anchor=south] at (3.3, -0.05) {$d$};
  \node[anchor=south] at (3.6, -0.065) {$d_{\textnormal{\tiny old}}$};

  \blockvector{50}
  \blockvector{60}
  \node[anchor=south] at (0.6, -0.05) {$R$};

  \blockvector{100}
  \blockvector{110}
  \node[anchor=south] at (1.1, -0.05) {$X$};

  \blockvector{150}
  \blockvector{160}
  \node[anchor=south] at (1.6, -0.05) {$Z$};

  \node[anchor=south] at (2.15, -0.05) {$\dots$};

  \blockvector{250}
  \blockvector{260}
  \node[anchor=south] at (2.65, -0.05) {$AP$};

  \end{scope}
\end{tikzpicture}

    \caption{The above figure displays the partitioning of all the vectors and intermediary
    Partition of vectors $R$, $X$, $Z$, $AZ$, $P$, and $AP$ along with
  $t\times t$ working arrays $c$, $d$, and $d_{\old}$, as in~\cref{ecg}, for $t=2$.}\label{fig:vector_partitioning}
\end{figure}
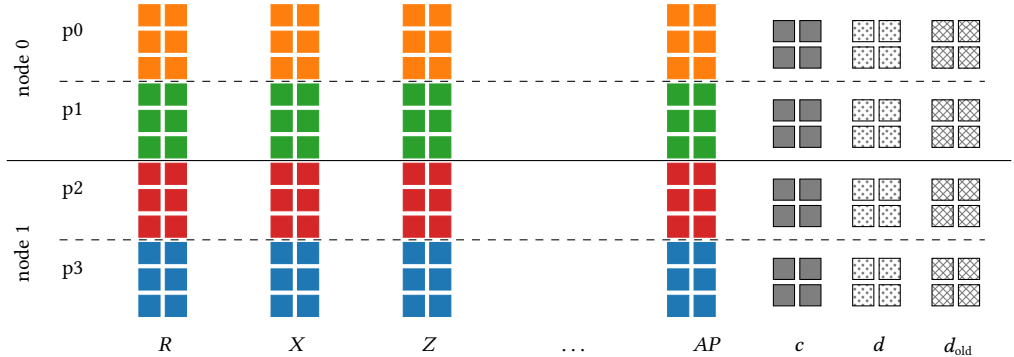

The scalability of a direct implementation of~\cref{ecg} is limited~\cite{scalable_enlarged},
however, this is improved
by fusing communication
and by executing the system solve in~\cref{alg:searchdir} in~\cref{ecg} on each process.
This is accomplished in~\cite{scalable_enlarged} by decomposing the computation of $P$
into several steps as described in~\cref{calc_pk}.
\begin{algorithm2e}[!ht]
  \DontPrintSemicolon%
  $AZ \leftarrow A * Z$\tcc*{sparse matrix-block vector multiplication}
  $Z^T AZ \leftarrow Z^T * AZ$\tcc*{$t$ inner products}
  $C^T C \leftarrow Z^T AZ$\tcc*{Cholesky factorization}\label{pk_chol}
  $P \leftarrow \text{solve}\,P * C = Z$\tcc*{Triangular solve with multiple right sides}\label{pk_line}
  \caption{Calculating $P := Z(Z^T A Z)^{-1/2}$}\label{calc_pk}
\end{algorithm2e}
The $t \times t$ product $Z^T (AZ)$ is
stored locally on every process in the storage space of
$c$, as shown in~\cref{fig:vector_partitioning}. The Cholesky factorization on~\cref{pk_chol} of~\cref{calc_pk}
is performed simultaneously on every process, yielding a (local) copy
of $C$.   Then each process performs a local triangular system solve using the local vector values of $Z$
to construct the portion of $P$ (see~\cref{pk_line}).
Similarly, an additional sparse matrix block vector product
$AP = A*P$ is avoided by noting that $AP$ is constructed using
\begin{equation*}
    AP \leftarrow \texttt{Triangular Solve with Multiple Right Sides of } AP * C = AZ
\end{equation*}
since the product $AZ$ and the previous iteration's $AP = A * P$ are already stored.

\cref{ecg_kernels} summarizes our implementation in terms of computational kernels,
with the on process computation in terms of floating point operations along with the associated
type of communication.
The remainder of the calculations within ECG consist
of local block vector updates, as well as block vector inner products for the
values $c$, $d$, and $d_{\old}$.
A straightforward approach is to compute these independently within the algorithm,
resulting in four \texttt{MPI\_Allreduce} global communications per iteration.
However, since the input data required to calculate $c$, $d$, and $d_{\old}$ are
available on~\cref{alg:c} in~\cref{ecg} when $c$ is computed, a single global reduction
is possible.
The implementation described
in~\cref{ecg_kernels} highlights a single call to
\texttt{MPI\_Allreduce} for all of these values and reducing them in the same buffer.
This reduces the number of \texttt{MPI\_Allreduce} calls to two per iteration.
\begin{algorithm2e}[!ht]
          \DontPrintSemicolon%
          SpMV\;
          Vector Initialization\;
          \For{$k = 1,\dots$}{%
SpMBV                        \atcp{$2 \cdot \frac{\nnz}{p} \cdot t$ \hspace{1em} Comm: point-to-point}
Block inner product          \atcp{$2 \cdot \frac{n}{p} \cdot t^2$ \hspace{1.5em} Comm: global all reduce}
Cholesky decomposition       \atcp{$\frac{1}{6} \cdot t^3$}
Triangular solves            \atcp{$2 \cdot \frac{1}{2} \cdot t^2$}
Block inner product          \atcp{$2 \cdot \frac{n}{p} \cdot t^2$ \hspace{1.5em} Comm: global all reduce}
Block vector addition        \atcp{$2 \cdot \frac{n}{p} \cdot t$}
Block vector \texttt{axpy}   \atcp{$2 \cdot \frac{n}{p} \cdot t$}
          }
          \caption{Enlarged Conjugate Gradient by Kernel}\label{ecg_kernels}
\end{algorithm2e}

\begin{figure}[!ht]
    \centering
    \includegraphics{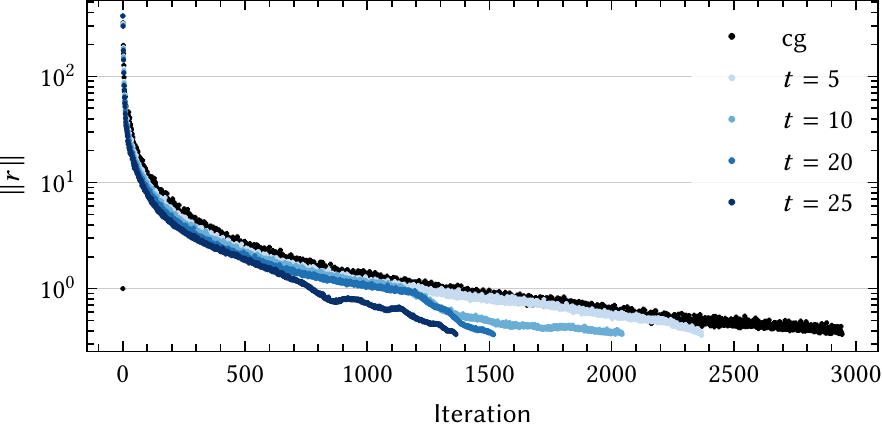}
    \caption{Residual history
      for CG and ECG with various enlarging factors $t$ for~\cref{ex:mfem}.
    }\label{fig:mfem_convergence}
\end{figure}

From~\cref{ecg_kernels}, we note that computation and communication
per iteration costs of ECG have increased over that of parallel CG\@. For
our implementation, the number of collective communication calls to
\texttt{MPI\_Allreduce} has remained the same as CG (at two),
but the number of values in the global reductions has increased from a single float
in each of CG's global reductions to $t^2$ and $3t^2$.
The singular SpMV from CG has increased to a sparse matrix block vector product (SpMBV),
which does not increase the number of point-to-point messages, but does increase the
size of the messages being communicated by the enlarging factor $t$.
Additionally, the local computation for \textit{each kernel} has increased by a factor of $t$.
ECG uses these extra per iteration requirements to reduce the total number of iterations
to convergence, resulting in fewer iterations than CG as seen in~\cref{fig:mfem_convergence}.

\subsection{Per Iteration Performance}

\begin{figure}[!htb]
    \centering
    \includegraphics[width=\textwidth]{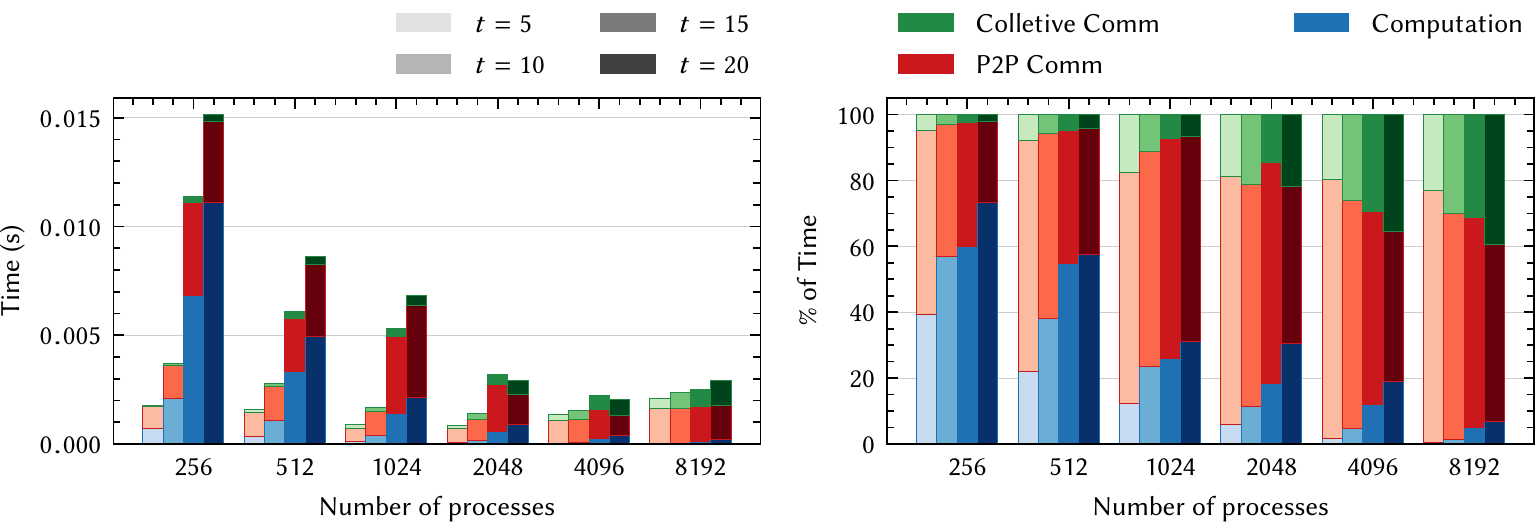}
    \caption{Time (left) and percentage of time (right) for a single
      iteration of ECG
      for various block vector sizes, $t$, and processor counts
    for~\cref{ex:mfem}.
    The three varying shades for each
    $t$ value represent the amount of overall time spent in on-process computation
    point-to-point communication, and collective communication.}\label{fig:ecg_profiling}
\end{figure}

\begin{figure}[!htb]
    \includegraphics[width=0.65\textwidth]{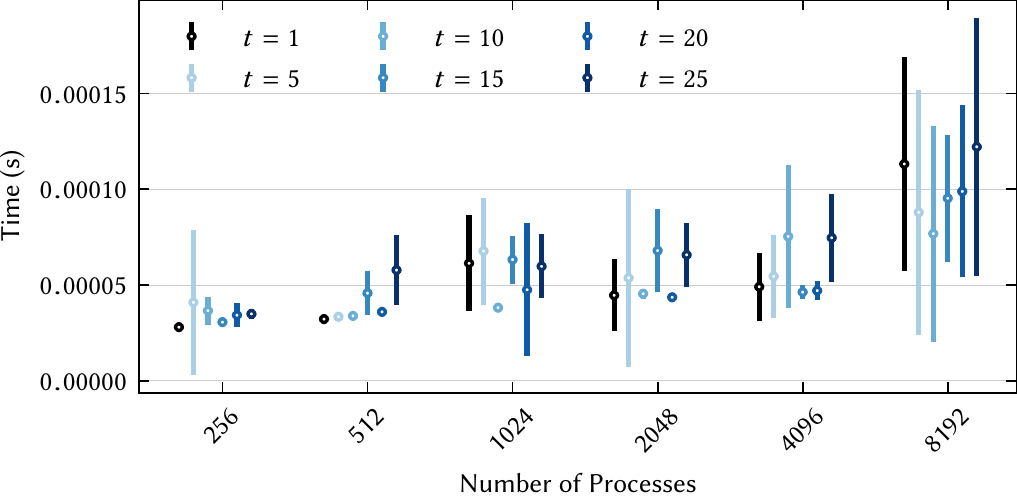}
    \caption{Time for a single inner product (right)
      for various block vector sizes, $t$, and processor counts for~\cref{ex:mfem}. Vertical lines denote the confidence intervals.}\label{fig:bvinner_profiling}
\end{figure}

In~\cref{fig:ecg_profiling} we decompose the performance of a single iteration of ECG
for~\cref{ex:mfem} into (local) computation, point-to-point communication, and collective
communication.
Performance tests were executed on Blue Waters~\cite{BW1,BW2}.
Each test is the average of 20 iterations of ECG\@; reported times are the maximum
average time recorded for any single process.
At small scales, local computation dominates performance, while at larger scales,
the point-to-point communication in the single SpMBV kernel and the collective
communication in the block vector inner products become the bottleneck in ECG\@.
\Cref{fig:ecg_profiling} also shows the time spent in a single inner product.  While
we observe growth with the number of processes, as expected, the relative
cost (and growth) within ECG remains low.
Importantly, increasing $t$ at high processor counts does not significantly contribute
to cost. This is shown in~\cref{fig:bvinner_profiling}, where the
mean runtime for various block vector inner products all fall within
each other's confidence intervals.
This suggests that increasing $t$ to drive down the iteration count will
have little affect on the per iteration cost of the two calls to \texttt{MPI\_Allreduce},
and in fact, will result in fewer total calls to it due to the reduction in iterations.

The remainder of this section focuses on accurately predicting the performance
of a single iteration of ECG through robust performance models.
In particularly, SpMBV communication is addressed in detail in~\cref{sec:block_nap} where we
discuss new node-aware communication techniques for blocked data.

\subsection{Performance Modeling}

To better understand the timing profiles in~\cref{fig:ecg_profiling}, we develop
performance models.
Below, we present two different models for the performance of communication within
a single iteration of ECG\@.
First, consider the standard postal model for communication, which
represents the maximum amount of time required for communication by an individual
process in a single iteration of ECG as
\begin{equation}\label{eq:ecg_comm_postal}
  T_{\text{postal}} = \underbrace{%
                             \alpha \cdot m + \frac{s \cdot t}{R_b}
                            }_\text{point to point} +
                 \underbrace{%
                             2\cdot\alpha \cdot \log(p) + \frac{f\cdot4\cdot t^2}{R_b}
                            }_\text{collective}
\end{equation}
where $f$ is the number of bytes for a floating point number~---~e.g.\ $f=8$.
See~\cref{tab:parameters} for a complete description of model parameters.
As discussed in~\cref{sec:node_aware_models}, this model presents a misleading
picture on the performance of ECG at scale, particularly on current supercomputer
architectures where SMP nodes encounter injection
bandwidth limits when sending inter-node messages.
\begin{figure}[!ht]
  \centering
\includegraphics[width=\textwidth]{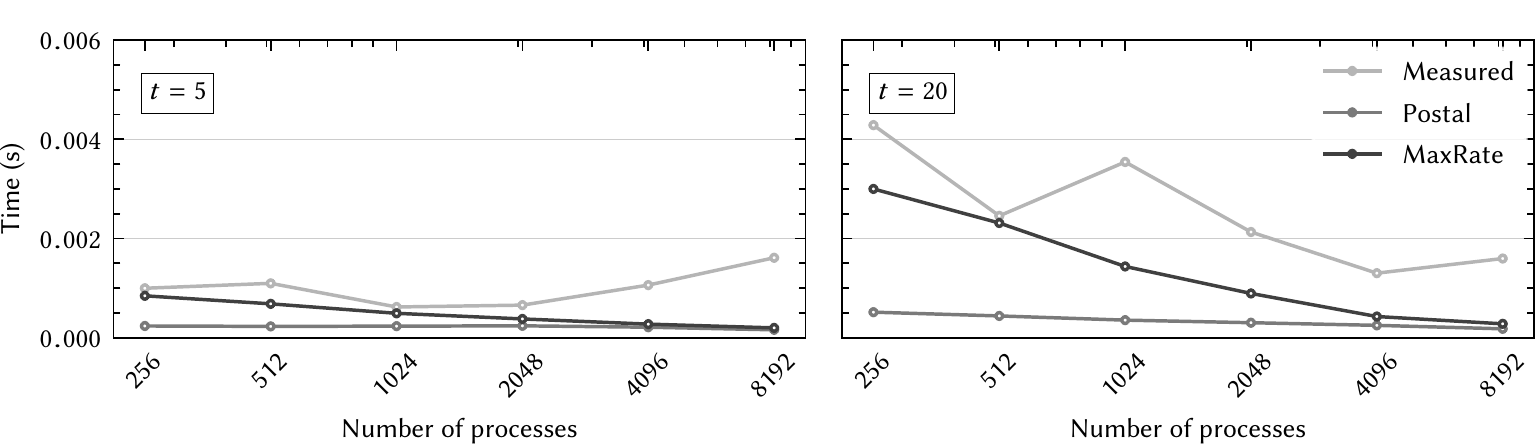}
\caption{Max-rate model versus the postal model for the point to point communication in one iteration of ECG for~\cref{ex:mfem} using various $t$. Measured runtimes are also included. (\textit{note:} $t=5$ and $t=20$ shown for brevity).}\label{fig:maxrate_proof}
\end{figure}

To improve the model
we drop in the max-rate model for the point-to-point
communication, resulting in
\begin{equation}\label{eq:ecg_comm_mr}
    T_{MR} = \underbrace{%
                         \alpha \cdot m +
                        \max \left( \frac{\ppn\cdot s\cdot t}{R_N}, \frac{s\cdot t}{R_b} \right)
                        }_\text{point to point}
           + \underbrace{%
                         2\cdot \alpha \cdot \log(p) + \frac{f\cdot4\cdot t^2}{R_b}
                        }_\text{collective}.
\end{equation}
\Cref{fig:maxrate_proof} shows that
the max-rate model provides a more accurate upper bound on the time spent in
point to point communication within ECG\@.
The term $2\cdot \alpha \cdot \log(p) + \frac{f\cdot4\cdot t^2}{R_b}$ remains unchanged in
\cref{eq:ecg_comm_postal,eq:ecg_comm_mr} to
represent the collective communication required for the two block
vector inner products. Each block vector inner product incurs latency
from requiring $\log(p)$ messages in an optimal implementation
of the \texttt{MPI\_Allreduce}.
More accurate models for modeling the communication of the \texttt{MPI\_Allreduce} in the
inner product exist, such as the logP model~\cite{logP} and logGP model~\cite{logGP},
but optimization of the reduction is outside the scope of this paper, so we leave
the postal model for representing the performance of the inner products.

Modeling the computation within an iteration of ECG is straightforward.
The computation for a single iteration of ECG is written as the sum
of the kernel floating point operations in~\cref{ecg_kernels},
which results in the following
\begin{equation}\label{eq:comp}
    T_{comp} = \gamma \cdot\left((2 + 2 t) \frac{\nnz}{p} + (4t + 4t^2) \frac{n}{p} +
                \frac{1}{2}t^2 + \frac{1}{6}t^3\right)
\end{equation}
where $\gamma$ is the time required to compute a single floating point operation.
In total, we arrive at the following model for a single iteration
of ECG
\begin{equation}\label{eq:ecg_iter}
    T_{ECG} = \alpha \cdot m +
        \max \left( \frac{\ppn\cdot s\cdot t}{R_N}, \frac{s\cdot t}{R_b} \right) +
        2\cdot\alpha \cdot \log(p) + \frac{f\cdot4\cdot t^2}{R_b}
                + \gamma \cdot\left((2 + 2 t) \frac{\nnz}{p}
                + (4t + 4t^2) \frac{n}{p}
                + \frac{1}{2}t^2
                + \frac{1}{6}t^3\right).
\end{equation}
Using this model, we can predict the reduction in the amount of
time spent in point-to-point communication using the multi-step
communication techniques presented in \cref{sec:node_aware_models}
for a single iteration of ECG
for~\cref{ex:mfem}~---~see~\cref{tab:ecg_model_percents}.
\begin{table}
\begin{subtable}{\textwidth}
\centering
  \begin{tabular}{c|cc|cc|cc|cc|cc|cc}
    \toprule
    \multicolumn{1}{c}{procs $\rightarrow$}    &
    \multicolumn{2}{c}{256} &
    \multicolumn{2}{c}{512} &
    \multicolumn{2}{c}{1024} &
    \multicolumn{2}{c}{2048} &
    \multicolumn{2}{c}{4096} &
    \multicolumn{2}{c}{8192}\\
    \multicolumn{1}{c}{$t$} &
    \scriptsize{m-s} & \multicolumn{1}{c}{\scriptsize{std}} &
    \scriptsize{m-s} & \multicolumn{1}{c}{\scriptsize{std}} &
    \scriptsize{m-s} & \multicolumn{1}{c}{\scriptsize{std}} &
    \scriptsize{m-s} & \multicolumn{1}{c}{\scriptsize{std}} &
    \scriptsize{m-s} & \multicolumn{1}{c}{\scriptsize{std}} &
    \scriptsize{m-s} & \multicolumn{1}{c}{\scriptsize{std}}\\
    \midrule
5  & \hlb{42.2} & {55.6} & \hlb{57.9} & {70.0} & \hlr{71.7} & {70.2} & \hlr{80.1} & {75.5} & \hlr{83.5} & {78.9} & \hlr{81.9} & {76.6} \\
10 & \hlb{21.2} & {40.0} & \hlb{34.5} & {56.2} & \hlb{51.2} & {65.2} &      65.9  &  67.5  & \hlr{75.9} & {69.1} & \hlr{77.7} & {68.7} \\
15 & \hlb{13.0} & {37.6} & \hlb{22.2} & {40.4} & \hlg{35.2} & {66.7} & \hlb{48.6} & {67.0} & \hlr{60.0} & {58.5} &      60.5  &  63.8  \\
20 & \hlb{9.3 } & {24.5} & \hlg{16.5} & {38.3} & \hlg{27.4} & {62.3} & \hlb{40.6} & {47.6} & \hlr{54.4} & {45.5} & \hlr{57.2} & {53.6} \\
    \bottomrule
  \end{tabular}
  \caption{2-Step Communication}\label{tab:ecg_percents_2step}
\end{subtable}

\bigskip
\begin{subtable}{\textwidth}
\centering
  \begin{tabular}{c|cc|cc|cc|cc|cc|cc}
    \toprule
    \multicolumn{1}{c}{procs $\rightarrow$}    &
    \multicolumn{2}{c}{256} &
    \multicolumn{2}{c}{512} &
    \multicolumn{2}{c}{1024} &
    \multicolumn{2}{c}{2048} &
    \multicolumn{2}{c}{4096} &
    \multicolumn{2}{c}{8192}\\
    \multicolumn{1}{c}{$t$} &
    \scriptsize{m-s} & \multicolumn{1}{c}{\scriptsize{std}} &
    \scriptsize{m-s} & \multicolumn{1}{c}{\scriptsize{std}} &
    \scriptsize{m-s} & \multicolumn{1}{c}{\scriptsize{std}} &
    \scriptsize{m-s} & \multicolumn{1}{c}{\scriptsize{std}} &
    \scriptsize{m-s} & \multicolumn{1}{c}{\scriptsize{std}} &
    \scriptsize{m-s} & \multicolumn{1}{c}{\scriptsize{std}}\\
    \midrule
    5  & \hlg{29.6} & {55.6} & \hlg{43.0} & {70.0} & \hlb{54.8} & {70.2} & \hlb{64.2} & {75.5} & \hlb{72.3} & {78.9} & \hlb{71.4} & {76.6} \\
    10 & \hlg{14.6} & {40.0} & \hlg{24.4} & {56.2} & \hlg{35.9} & {65.2} & \hlb{49.5} & {67.5} & 65.6 & 69.1 & 68.5 & 68.7 \\
    15 & \hlg{9.2} & {37.6} & \hlg{15.8} & {40.4} & \hlg{23.9} & {66.7} & \hlg{34.4} & {67.0} & \hlb{49.9} & {58.5} & \hlb{50.8} & {63.8} \\
    20 & \hlb{6.7} & {24.5} & \hlg{11.9} & {38.3} & \hlg{18.6} & {62.3} & \hlb{28.7} & {47.6} & \hlr{45.9} & {45.5} & 48.8 & 53.6 \\
    \bottomrule
  \end{tabular}
  \caption{3-Step Communication}\label{tab:ecg_percents_3step}
\end{subtable}
\caption{Modeled percentage of time to be spent in point-to-point communication for
    multistep (``m-s'') compared against the measured percentage of time spent
    in point-to-point communication for standard (``std'') in a single iteration of ECG
    for \cref{ex:mfem} with varying $t$ and processor counts on Blue Waters.
    Blue values correspond to values for which the percentage of time
    \hlc[hlblue]{decreased by $5\% - 20\%$},
    green values are a \hlc[hlgreen]{decrease of over $20\%$}, and
    red values predict an \hlc[hlred]{increase over the original percentage}.
  }\label{tab:ecg_model_percents}
\end{table}

We see that ECG is still limited at
large processor counts even when substituting the node-aware
communication techniques to replace the costly point-to-point
communication observed in~\cref{fig:ecg_profiling}.
The models do predict a large amount of speedup for most cases,
however, when using 3-step communication, suggesting that
node-aware communication techniques can reduce the large
point-to-point bottleneck observed in the performance study.
While a large communication cost stems equally from the
collective communication the \texttt{MPI\_Allreduce} operations,
their performance is dependent upon underlying \texttt{MPI}
implementation and outside the scope of this paper.
We address the point-to-point communication performance further
in~\cref{sec:block_nap} by analyzing it through the lens of
node-aware communication techniques, optimizing them to achieve
the best possible performance at scale.

\section{Optimized Communication for Blocked Data}\label{sec:block_nap}

As discussed in~\cref{sec:performance}, scalability for ECG
is limited by the sparse matrix-block vector multiplication (SpMBV) kernel defined as
\begin{equation}\label{eq:spmbv}
   A \cdot V \rightarrow W,
\end{equation}
with $A \in \Rnn$ and $V$, $V \in \Rnt$, where $1 < t \ll n$.
Due to the block vector structure of $X$, each message in
a SpMV is increased by a factor of $t$ (see~\cref{fig:node_aware_spmbv}).
The larger messages associated with $t>1$ increase the amount of time spent in the point-to-point
communication at larger scales, making the SpMBV operation an ideal candidate for node-aware messaging approaches.
\begin{figure}
  \small %
  \definecolor{tab-blue}{HTML}{1f77b4}
\definecolor{tab-orange}{HTML}{ff7f0e}
\definecolor{tab-green}{HTML}{2ca02c}
\definecolor{tab-red}{HTML}{d62728}
\definecolor{tab-purple}{HTML}{9467bd}
\definecolor{tab-brown}{HTML}{8c564b}
\definecolor{tab-pink}{HTML}{e377c2}
\definecolor{tab-gray}{HTML}{7f7f7f}
\definecolor{tab-olive}{HTML}{bcbd22}
\definecolor{tab-cyan}{HTML}{17becf}

\newcommand{\aij}[5]
{
  \draw[fill=#4!#5,draw=#4] (#2/10+3/10, 1/10+#3/10-#1/10+0.01) rectangle +(0.08, 0.08);
}

\newcommand{\onevec}[1]
{
  \begin{scope}[shift={(#1,0)}]
  \aij{1}{1}{12}{tab-orange}{100}
  \aij{2}{1}{12}{tab-orange}{100}
  \aij{3}{1}{12}{tab-orange}{100}
  \aij{4}{1}{12}{tab-green}{100}
  \aij{5}{1}{12}{tab-green}{100}
  \aij{6}{1}{12}{tab-green}{100}
  \aij{7}{1}{12}{tab-red}{100}
  \aij{8}{1}{12}{tab-red}{100}
  \aij{9}{1}{12}{tab-red}{100}
  \aij{10}{1}{12}{tab-blue}{100}
  \aij{11}{1}{12}{tab-blue}{100}
  \aij{12}{1}{12}{tab-blue}{100}
  \end{scope}
}

\begin{tikzpicture}[x=100pt,y=100pt]
  \begin{scope}[shift={(5pt, 15pt)}]

  \foreach \z in {1, ..., 12}{%
    \foreach \y in {1,...,12}{%
      \def\x{\z/10+3/10}
      \draw[draw=tab-gray!20] (\x, \y/10+0.01) rectangle +(0.08, 0.08);
    }
  }

  \aij{1}{1}{12}{tab-orange}{100}
  \aij{2}{2}{12}{tab-orange}{100}
  \aij{3}{3}{12}{tab-orange}{100}
  \aij{1}{3}{12}{tab-orange}{100}
  \aij{3}{2}{12}{tab-orange}{100}

  \aij{4}{4}{12}{tab-green}{100}
  \aij{5}{5}{12}{tab-green}{100}
  \aij{6}{6}{12}{tab-green}{100}
  \aij{4}{5}{12}{tab-green}{100}
  \aij{5}{4}{12}{tab-green}{100}
  \aij{5}{6}{12}{tab-green}{100}

  \aij{7}{7}{12}{tab-red}{100}
  \aij{8}{8}{12}{tab-red}{100}
  \aij{9}{9}{12}{tab-red}{100}
  \aij{9}{7}{12}{tab-red}{100}

  \aij{10}{10}{12}{tab-blue}{100}
  \aij{11}{11}{12}{tab-blue}{100}
  \aij{12}{12}{12}{tab-blue}{100}
  \aij{10}{11}{12}{tab-blue}{100}
  \aij{12}{10}{12}{tab-blue}{100}

  \aij{1}{6}{12}{tab-orange}{20}
  \aij{2}{4}{12}{tab-orange}{20}
  \aij{2}{5}{12}{tab-orange}{20}
  \aij{3}{5}{12}{tab-orange}{20}

  \aij{4}{1}{12}{tab-green}{20}
  \aij{5}{2}{12}{tab-green}{20}
  \aij{6}{1}{12}{tab-green}{20}

  \aij{7}{11}{12}{tab-red}{20}
  \aij{8}{12}{12}{tab-red}{20}
  \aij{9}{10}{12}{tab-red}{20}

  \aij{11}{9}{12}{tab-blue}{20}
  \aij{12}{8}{12}{tab-blue}{20}

  \aij{1}{10}{12}{tab-orange}{0}
  \aij{2}{8}{12}{tab-orange}{0}
  \aij{2}{9}{12}{tab-orange}{0}
  \aij{2}{12}{12}{tab-orange}{0}
  \aij{3}{7}{12}{tab-orange}{0}
  \aij{3}{10}{12}{tab-orange}{0}

  \aij{4}{9}{12}{tab-green}{0}
  \aij{5}{11}{12}{tab-green}{0}
  \aij{6}{9}{12}{tab-green}{0}

  \aij{7}{2}{12}{tab-red}{0}
  \aij{8}{4}{12}{tab-red}{0}
  \aij{9}{3}{12}{tab-red}{0}

  \aij{10}{5}{12}{tab-blue}{0}
  \aij{11}{4}{12}{tab-blue}{0}
  \aij{12}{1}{12}{tab-blue}{0}

  \onevec{1.3}
  \onevec{1.4}
  \onevec{1.5}

  \onevec{1.8}
  \onevec{1.9}
  \onevec{2.0}

  \node[below, align=center] (a) at (1.0, 0.05) {$A$};
  \node[below, align=center, right= 0.5 of a] (b) {$*$};
  \node[below, align=center, right= .08 of b] (c) {$V$};
  \node[below, align=center, right= .1 of c] (d) {$\rightarrow$};
  \node[below, align=center, right= .05 of d] (e) {$W$};

  \draw[black, solid] (0, 0.70) -- (2.5, .70);
  \node[rotate=90,anchor=north] at (0.0, 1.05) {node $0$};
  \node[rotate=90,anchor=north] at (0.0, .35) {node $1$};

  \draw[black, dashed] (0.2, 1.0) -- (2.4, 1.0);
  \draw[black, dashed] (0.2, 0.4) -- (2.4, 0.4);
  \node[anchor=north] at (0.25, 1.25) {p0};
  \node[anchor=north] at (0.25, 0.95) {p1};
  \node[anchor=north] at (0.25, 0.65) {p2};
  \node[anchor=north] at (0.25, 0.35) {p3};

  \end{scope}
\end{tikzpicture}

  \caption{Sparse matrix-block vector multiplication (cf.~\cref{fig:spmv_nodeaware}).}\label{fig:node_aware_spmbv}
\end{figure}

\subsection{Performance Modeling}\label{sec:blockmodel}

Recalling the node-aware communication models from~\cref{sec:node_aware_models},
we augment the 2-step and 3-step models with block vector size, $t$.  As a result,
the 2-step model in~\cref{eq:model2} becomes
\begin{equation}\label{eq:model2_bv}
  T_{total} = \underbrace{%
                          \alpha \cdot \nproctonode +
                          \max\left(%
                            \frac{t\cdot\snode}{R_N},
                            \frac{t\cdot\sproc}{R_b}
                          \right)
                         }_{\text{inter-node}}
             + \underbrace{%
                           \alpha_{\ell} \cdot (\ppn - 1) +
                           t \cdot \frac{\sproc}{R_{b,\ell}}
                          }_{\text{intra-node}}
\end{equation}
while the 3-step model in~\cref{eq:model3} becomes
\begin{equation}\label{eq:model3_bv}
  T_{total} =  \underbrace{%
                           \alpha \cdot \frac{\nnodetonode}{\ppn} +
                           \max\left(%
                             \frac{t\cdot\snode}{R_N},
                             \frac{t\cdot\sproc}{R_b}
                           \right)
                          }_{\text{inter-node}}
             + \underbrace{%
                           2\cdot\left(%
                             \alpha_{\ell} \cdot (\ppn - 1) +
                             t \cdot \frac{\snodetonode}{R_{b,\ell}}
                           \right)
                          }_{\text{intra-node}}.
\end{equation}
In both models, $t$ impacts maximum rate, a term that
is relatively small for $t=1$ and
dominates the inter-node portion of the communication as $t$ grows.
Since messages are increased by a factor of $t$, the single buffer
used in 3-step communication quickly reaches the network injection bandwidth limits.
Using multiple buffers, as in 2-step communication, helps mitigate the issue,
however more severe imbalance persists
since the amount of data sent to different nodes is often widely varying.

\begin{figure}[!ht]
  \centering
  \includegraphics[width=\textwidth]{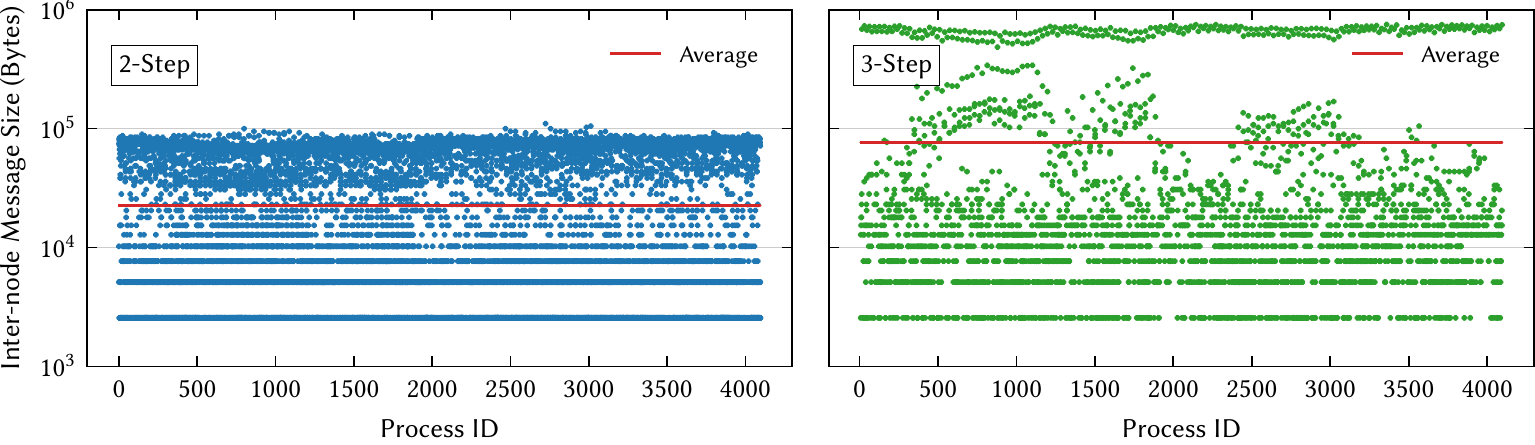}
  \caption{The inter-node message sizes across all processes for 2-Step
            and 3-step communication when performing the SpMBV
            for~\cref{ex:mfem} when $t=20$ across 4096 processes on Blue Waters.
            The average message size across all processes is marked by a red line.}\label{fig:msg_sizes_p4096}
\end{figure}

\Cref{fig:msg_sizes_p4096} shows every inter-node message sent by a
single process alongside the message size for~\cref{ex:mfem} when performing the SpMBV
kernel with 4096 processes and 16 processes per node.
We present the message sizes for 3-step and 2-step communication, noting that the
overall number of inter-node messages decreases when using 3-step
communication, but the average message size sent by a single process increases.
For $t=20$, the maximum message size nears $10^6$ for 3-step communication, while
the maximum message size barely reaches $10^5$ for 2-step communication. Additionally,
there is clear imbalance in the inter-node message sizes for 2-step communication
with messages ranging in size from $10^3$--$10^5$ Bytes.

\subsection{Profiling}

We next apply the node-aware communication strategies
presented for SpMVs and SpGEMMs in~\cref{sec:node_aware_comm} to the
SpMBV kernel within ECG\@.

\begin{figure}[!ht]
  \centering
\includegraphics[width=\textwidth]{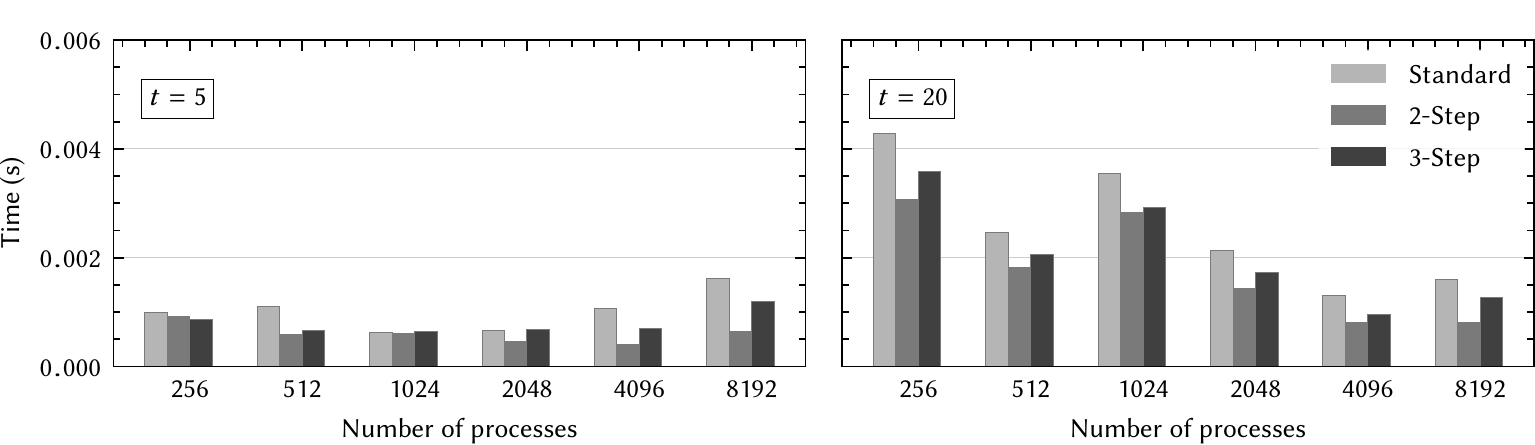}
\caption{Strong scaling results for~\cref{ex:mfem} using standard, 2-step, and 3-step
communication in the SpMBV (\textit{note:} $t=5$ and $t=20$ shown for brevity).}\label{fig:mfem_spmbv_comparison}
\end{figure}

\cref{fig:mfem_spmbv_comparison} displays the performance of standard, 2-step, and 3-step
communication when applied to the SpMBV kernel for \cref{ex:mfem} with $t=5$ and
$t=20$.
The 2-step communication appears outperform in most cases. This is due to the
large amount of data to be sent off-node that is split across many processes. We see 2-step
communication performing better due to the term $\alpha \cdot \nproctonode$
in~\cref{eq:model2_bv} being smaller than the $\alpha \cdot \nnodetonode$ term
in the 3-step communication model~\cref{eq:model3_bv} due to multiplication by the
factor $t$. In fact, all counts measured in the ECG profiling section are now multiplied
by the enlarging factor $t$. While the traditional SpMV shows speedup with 3-step
communication, we now see that 2-step is generally the best fit for our methods as message size, and thus $t$, increases.

Next we consider node-aware communication performance results for a
subset of the largest matrices in the SuiteSparse matrix
collection~\cite{ufl_matrices} (matrix details can be found
in~\cref{tab:matrices}).  These were selected based on size and density
to provide a variety of scenarios.
\begin{table}
  \begin{tabular}{lcccc}
    \toprule
    matrix      & rows/ cols      & nnz               & nnz/row  & density \\
    \midrule
    audikw\_1   & \num{943 695}   & \num{77 651 847}  & 82.3     & 8.72e-5  \\
    Geo\_1438   & \num{1 437 960} & \num{60 236 322}  & 41.9     & 2.91e-05  \\
    bone010     & \num{986 703}   & \num{47 851 783}  & 48.5     & 4.92e-05  \\
    Emilia\_923 & \num{923 136}   & \num{40 373 538}  & 43.7     & 4.74e-05  \\
    Flan\_1565  & \num{1 565 794} & \num{114 165 372} & 72.9     & 4.66e-05  \\
    Hook\_1498  & \num{1 498 023} & \num{59 374 451}  & 39.6     & 2.65e-05  \\
    ldoor       & \num{952 203}   & \num{42 493 817}  & 44.6     & 4.69e-05  \\
    Serena      & \num{1 391 349} & \num{64 131 971}  & 46.1     & 3.31e-05  \\
    thermal2    & \num{1 228 045} & \num{8 580 313}   & 7.0      & 5.69e-06  \\
    \bottomrule
  \end{tabular}
  \caption{Test Matrices.}\label{tab:matrices}
\end{table}

\begin{figure}[!ht]
  \centering
\includegraphics[width=1.0\textwidth]{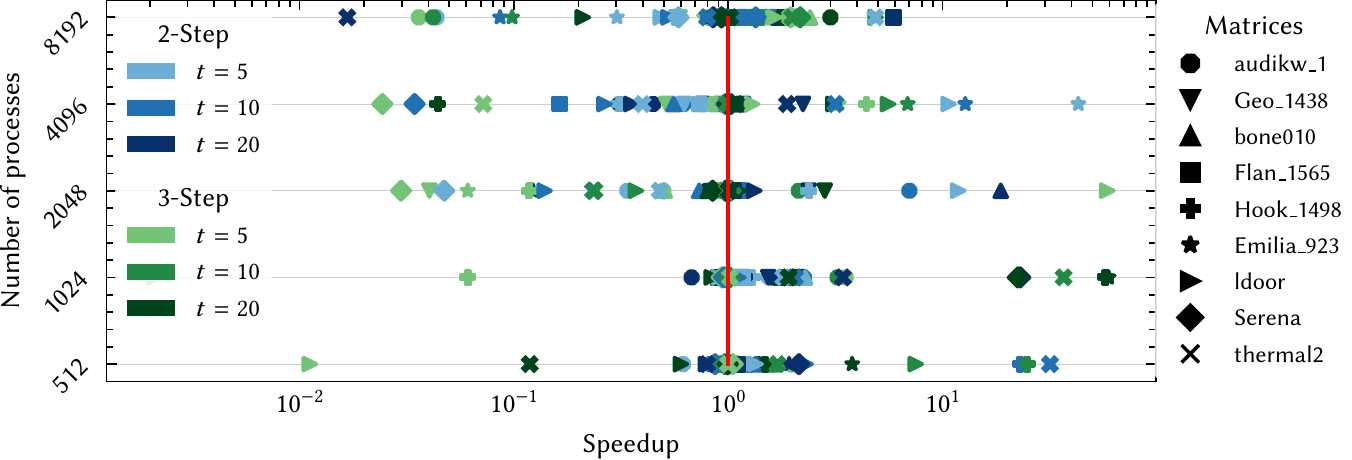}
\caption{Speedup of 2-step and 3-step communication over standard communication
in the SpMBV for various matrices from the SuiteSparse matrix collection on Blue Waters.
The red line marks 1.0, or no speedup.}\label{fig:bw_spmbv_comparison}
\end{figure}

\begin{figure}[!ht]
  \centering
\includegraphics[width=1.0\textwidth]{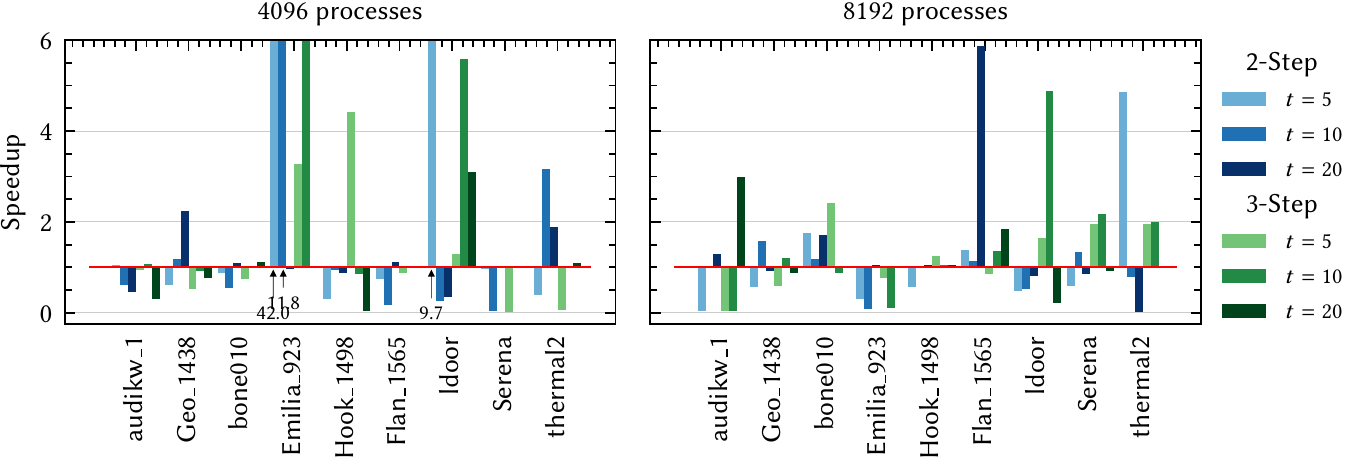}
\caption{Speedup of 2-step and 3-step communication over standard communication
in the SpMBV for various matrices on Blue Waters
for 4096 and 8192 processes. The red line marks 1.0, or no speedup. Speedups greater
than 6.0 are given via annotations on the plot.}\label{fig:bw_spmbv_comparison_zoom}
\end{figure}

While 2-step communication is effective in many instances, it is not always the most optimal
communication strategy, as depicted in \cref{fig:bw_spmbv_comparison}.
Unlike the results for~\cref{ex:mfem}, 3-step
and 2-step communication do not always outperform standard
communication, and in some instances (4096 processes), for most
values of $t$ there is performance degradation. This is seen
more clearly in \cref{fig:bw_spmbv_comparison_zoom}.

It is important to highlight cases where only a single node-aware communication technique
results in performance deterioration over standard communication. Distinct examples include
\texttt{Geo\_1438} and \texttt{thermal2} on 4096 processes. Both of these matrices benefit from
2-step communication for $t=10$ and $20$, but performance degrades when
using 3-step communication. Another example is the performance of \texttt{ldoor}
on 8192 processes (seen in \cref{fig:bw_spmbv_comparison_zoom}).
For $t=5$ and $10$, \texttt{ldoor} results in performance degradation using
2-step communication, but up to $5\times$ speedup over standard communication
when using 3-step communication.
These cases highlight that while one node-aware
communication technique underperforms in comparison to standard communication, the other node
aware technique is still much faster. Using this as the key motivating factor,
we discuss an optimal
node-aware communication technique for blocked data in~\cref{sec:opt_comm}.

\subsection{Optimal Communication for Blocked Data}\label{sec:opt_comm}

When designing an optimal communication scheme for the blocked data format, the main
consideration is the impact the number of vectors within the block have on the size
of messages communicated. The effects message size and message number have on
performance can vary based on machine, hence we present results for Blue Waters
alongside Lassen~\cite{lassen}. Lassen, a 23-petaflop IBM system at Lawrence
Livermore National Laboratory, consists of 792 nodes connected via a Mellanox
100 Gb/s Enhanced Data Rate (EDR) InfiniBand network. Each node on Lassen
is dual-socket with 44 IBM Power9 cores and 4 NVIDIA Volta GPUs (which are
unused in our tests.)

\begin{figure}[!ht]
  \centering
  \includegraphics[width=\textwidth]{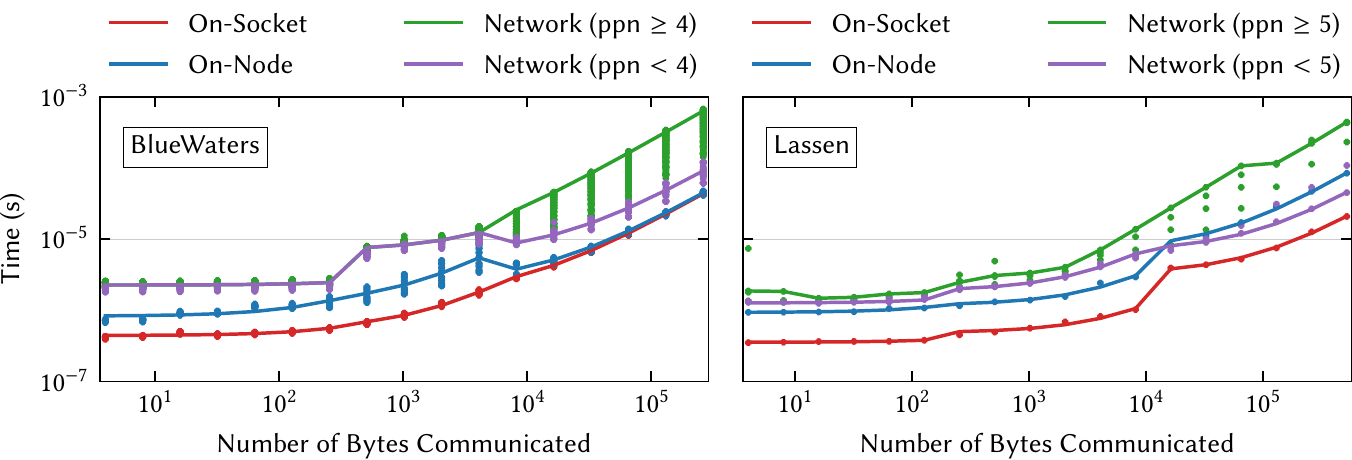}
  \caption{
  Communication time on-socket, on-node, or across the network on Blue Waters (left) and Lassen (right).
  Measured times are displayed as dots; solid lines represent
  max-rate model~\cite{maxrate_model}.  \ppn\ represents the number of processors participating in communication.
  (The Blue Waters data was initially presented in~\cite{Bienz_model} and is replotted here.)
  }\label{fig:comm_split}
\end{figure}

In~\cref{fig:comm_split}, we view the effects placement
of data and amount of data being communicated has on performance times for two different machines.
This figure shows the amount of time required to send various numbers
of bytes between two processes when they are located on the same node but
different sockets (blue) and on the same node and same socket (red) for the
machines Blue Waters (left) and Lassen (right). It also shows the amount of time
required to communicate between two processes on separate nodes when there are
less than 4 (Blue Waters) or 5 (Lassen) active processes communicating through the
network at the same time or greater than 4 or 5 processes communicating through the network
at the same time.
We see that as the number of bytes communicated
between two processes increases, it becomes increasingly important whether those
two processes are located on the same socket, node, or require communication through
the network.
For both machines, inter-node communication is fastest when message sizes are small, and there are few
messages being injected into the network.
On Blue Waters, intra-node communication is the fastest,
with the time being dependent on how physically
close the processes are located. For instance, when two processes are on the same socket,
communication is faster than when they are on the same node, but different sockets.
This is true for Lassen, as well, but cross-socket intra-node communication is not always faster than communicating through
the network. Once message sizes exceed $10^4$ bytes, and there are fewer than five
processes actively communicating, inter-node communication is faster than
two cross-socket intra-node processes communicating.
For both machines, however, we see that once a large enough communication volume is
reached, it becomes faster to split the inter-node data being sent across a subset of
the processes on a single node due to network contention as observed in~\cite{Bienz_model}.

\begin{figure}[!ht]
  \centering
  \includegraphics[width=\textwidth]{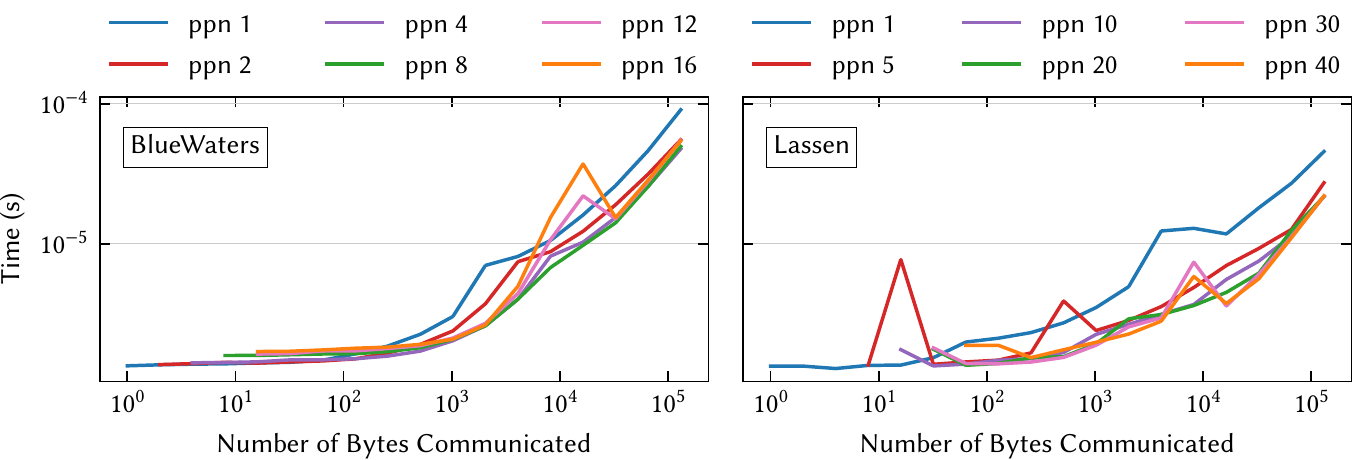}
  \caption{
    Time required to send data between two processes on separate nodes with the data
    communication split across \ppn\ processes. Timings for Blue Waters and
    Lassen are presented in the left and right plots, respectively.
   }\label{fig:node_pong}
\end{figure}

In addition to the importance of the placement of two communicating processes, the
total message volume and number of actively communicating processes plays a key
role in communication performance.
While it is extremely costly for every process on a node to send $10^5$ bytes
as seen in \cref{fig:comm_split},
there are performance benefits when splitting a large communication
volume across all processes on a node, depicted in \cref{fig:node_pong}.
Blue Waters sees modest performance benefits when splitting large messages
across multiple processes, whereas Lassen sees much greater performance benefits.

These observations help justify why 3-step communication would outperform 2-step
communication, and vice versa in
certain cases of the SpMBV profiling presented in~\cref{fig:maxrate_proof}.
Sending all messages in a single
buffer becomes impractical when the block size, $t$, is very large, but having each process
communicate with a paired process also poses problems when some of the inter-node
messages being sent are still very large, as seen in \cref{fig:msg_sizes_p4096}.

Motivated by the results above, we introduce an optimized multi-step communication process that combines the aspects of both the 3-step and 2-step communication techniques, and
using 3-step or 2-step communication when necessary. We reduce the number of
inter-node messages and conglomerate messages to be sent off node for certain cases when
the message sizes to be sent off-node are below a given threshold, and we
split the messages to be sent off-node across multiple processes when the message size is
larger than a given threshold. Hence, each node is determining the most optimal way
to perform its inter-node communication.
As a result, this nodal optimal communication eliminates the redundancies of data being
injected into the network, just as 3-step and 2-step do, but in some cases, it does not reduce  the number of inter-node messages as much as 3-step, and in fact can increase the number
of inter-node messages injected by a single node to be larger than those injected by 2-step, but never exceeding the total number of active processes per node.

The number of inter-node messages
sent can be represented by the following inequality,
\begin{equation}\label{eq:opt_messages}
  \nnodetonode \leq n_{opt} \leq \max(\nproctonode,\ppn)
\end{equation}
where $n_{opt}$ is the number of inter-node messages injected by a single process for the optimal node-aware communication,
and it is bounded below by the worst-case number of messages communicated
in 3-step communication \cref{eq:model3_bv} and above by the worst-case number
of messages communicated in 2-step communication \cref{eq:model2_bv} or \ppn, whichever
is greater.

\begin{figure}[!htp]
  \centering
  \small %
\definecolor{tab-blue}{HTML}{1f77b4}
\definecolor{tab-orange}{HTML}{ff7f0e}
\definecolor{tab-green}{HTML}{2ca02c}
\definecolor{tab-red}{HTML}{d62728}
\definecolor{tab-purple}{HTML}{9467bd}
\definecolor{tab-brown}{HTML}{8c564b}
\definecolor{tab-pink}{HTML}{e377c2}
\definecolor{tab-gray}{HTML}{7f7f7f}
\definecolor{tab-olive}{HTML}{bcbd22}
\definecolor{tab-cyan}{HTML}{17becf}

\newcommand{\anode}
{
  \begin{scope}[x=40pt,y=40pt]
  \draw[draw=tab-gray!80!black, fill=tab-gray!30] (0,     0) rectangle +(2, 2);
  \draw[draw=tab-orange!80!black,fill=tab-orange!25]  (0.5, 1.5) circle (0.35 and .45);
  \draw[draw=tab-green!80!black, fill=tab-green!25]   (1.5, 1.5) circle (0.35 and .45);
  \draw[draw=tab-red!80!black,   fill=tab-red!25]     (0.5, 0.5) circle (0.35 and .45);
  \draw[draw=tab-blue!80!black,  fill=tab-blue!25]    (1.5, 0.5) circle (0.35 and .45);
  \coordinate (p0) at (0.5, 1.4);
  \coordinate (p1) at (1.5, 1.4);
  \coordinate (p2) at (0.5, 0.4);
  \coordinate (p3) at (1.5, 0.4);
  \node[anchor=center] at (p0) {\scriptsize P0};
  \node[anchor=center] at (p1) {\scriptsize P1};
  \node[anchor=center] at (p2) {\scriptsize P2};
  \node[anchor=center] at (p3) {\scriptsize P3};
  \node[anchor=south] at (1.0, 2.0) {Node 0};

  \begin{scope}[shift={(90pt,0pt)}]
  \draw[dashed,draw=tab-gray!80!black, fill=tab-gray!30] (0,     0) rectangle +(2, 2);
  \draw[dashed,draw=tab-orange!80!black,fill=tab-orange!25]  (0.5, 1.5) circle (0.35 and .45);
  \draw[dashed,draw=tab-green!80!black, fill=tab-green!25]   (1.5, 1.5) circle (0.35 and .45);
  \draw[dashed,draw=tab-red!80!black,   fill=tab-red!25]     (0.5, 0.5) circle (0.35 and .45);
  \draw[dashed,draw=tab-blue!80!black,  fill=tab-blue!25]    (1.5, 0.5) circle (0.35 and .45);
  \coordinate (p4) at (0.5, 1.4);
  \coordinate (p5) at (1.5, 1.4);
  \coordinate (p6) at (0.5, 0.4);
  \coordinate (p7) at (1.5, 0.4);
  \node[anchor=center] at (p4) {\scriptsize P4};
  \node[anchor=center] at (p5) {\scriptsize P5};
  \node[anchor=center] at (p6) {\scriptsize P6};
  \node[anchor=center] at (p7) {\scriptsize P7};
  \node[anchor=south] at (1.0, 2.0) {Node 1};
  \end{scope}
  \end{scope}
}

\tikzset{
  doublearrowp/.style args={#1 and #2}{
    round cap-latex,line width=1pt,#1,shorten <=5pt,
    postaction={draw,round cap-latex,#2,line width=1pt/3,shorten <=5.5pt,shorten >=1.5pt},
  },
  doublearrowq/.style args={#1 and #2}{
    round cap-latex,line width=1pt,#1, shorten <=2pt, shorten >=5pt,
    postaction={draw,round cap-latex,#2,line width=1pt/3,shorten <=2.5pt,shorten >=6.5pt},
  },
  doublearrowz/.style args={#1 and #2}{
    round cap-latex,line width=1pt,#1,shorten <=2pt, shorten >=2pt,
    postaction={draw,round cap-latex,#2,line width=1pt/3,shorten <=2.5pt,shorten >=3.5pt},
  },
  doublearrowp2/.style args={#1}{
    round cap-latex,line width=1pt,#1,shorten <=5pt, shorten >=5pt,
  },
  doublearrowq2/.style args={#1}{
    round cap-latex,line width=1pt,#1, shorten <=2pt, shorten >=5pt,
  },
  doublearrowz2/.style args={#1}{
    round cap-latex,line width=1pt,#1,shorten <=5pt, shorten >=5pt,
  },
}
\begin{tikzpicture}[x=80pt,y=80pt]
  \begin{scope}[shift={(5pt, 5pt)}]
  \newcommand{\drawdata}[3]{
    \ifthenelse{\equal{#3}{above}}
    {
     \pgfmathsetmacro\zz{5pt}
    }
    {
     \pgfmathsetmacro\zz{-8pt}
    }

    \pgfmathsetmacro\m{#1-1}
    \foreach \j in {0,...,\m}{
      \draw[color=black!70, fill=tab-gray] ([xshift=-\m*2+4*\j pt,yshift=\zz]#2) rectangle +(0.04, 0.04);
    }
  }
  \begin{scope}[xshift=20pt,yshift=230pt]
    \fill[tab-gray!10] (-0.1,-0.1) rectangle +(2.35,2.5);
    \anode
    \draw[doublearrowp2=black] (p1) to (p0);
    \draw[doublearrowp2=black] (p3) to (p1);
    \drawdata{1}{p0}{above}
    \drawdata{2}{p1}{above}
    \drawdata{1}{p2}{below}
    \drawdata{3}{p3}{below}
    \begin{scope}[xshift=0pt,yshift=100pt]
    \anode
    \draw[doublearrowp2=black] (p5) to (p4);
    \draw[doublearrowp2=black] (p7) to (p4);
    \drawdata{1}{p4}{above}
    \drawdata{1}{p5}{above}
    \drawdata{1}{p7}{below}
    \end{scope}
    \node[rotate=90] at (-14pt,95pt) {Step 1};
  \end{scope}
  \begin{scope}[xshift=60pt,yshift=120pt]
    \node[rotate=90] at (-14pt,40pt) {Step 2};
    \anode
    \draw[doublearrowq2=black, <->, >=latex] ([yshift=8pt]p0) to[looseness=0.7,out=60, in=120] ([yshift=8pt]p4);

    \draw[doublearrowq2=black, ->, >=latex]  ([yshift=-8pt]p2) to[looseness=0.7,out=-60, in=-120] ([yshift=-8pt]p6);
    \draw[doublearrowq2=black, ->, >=latex] ([yshift=-8pt]p1) to[looseness=0.9,out=-60, in=-120] ([yshift=-8pt]p5);

    \drawdata{3}{p0}{above}
    \drawdata{3}{p1}{above}
    \drawdata{1}{p2}{below}

    \drawdata{3}{p4}{above}
  \end{scope}
  \begin{scope}[xshift=120pt,yshift=0pt]
    \node[rotate=90] at (-14pt,40pt) {Step 3};
    \anode
    \draw[doublearrowz2=black, <->, >=latex] (p0) -- (p1);
    \draw[doublearrowz2=black, <->, >=latex] (p1) -- (p3);
    \draw[doublearrowz2=black, <->, >=latex] (p3) -- (p2);
    \draw[doublearrowz2=black, <->, >=latex] (p2) -- (p0);
    \draw[doublearrowz2=black, <->, >=latex] (p0) -- (p3);
    \draw[doublearrowz2=black, <->, >=latex] (p2) -- (p1);

    \draw[doublearrowz2=black, <->, >=latex] (p4) -- (p5);
    \draw[doublearrowz2=black, <->, >=latex] (p5) -- (p7);
    \draw[doublearrowz2=black, <->, >=latex] (p7) -- (p6);
    \draw[doublearrowz2=black, <->, >=latex] (p6) -- (p4);
    \draw[doublearrowz2=black, <->, >=latex] (p4) -- (p7);
    \draw[doublearrowz2=black, <->, >=latex] (p5) -- (p6);
  \end{scope}

  \end{scope}
\end{tikzpicture}

  \caption{Node-aware communication on two nodes with four processes each.}\label{fig:optstep}
\end{figure}
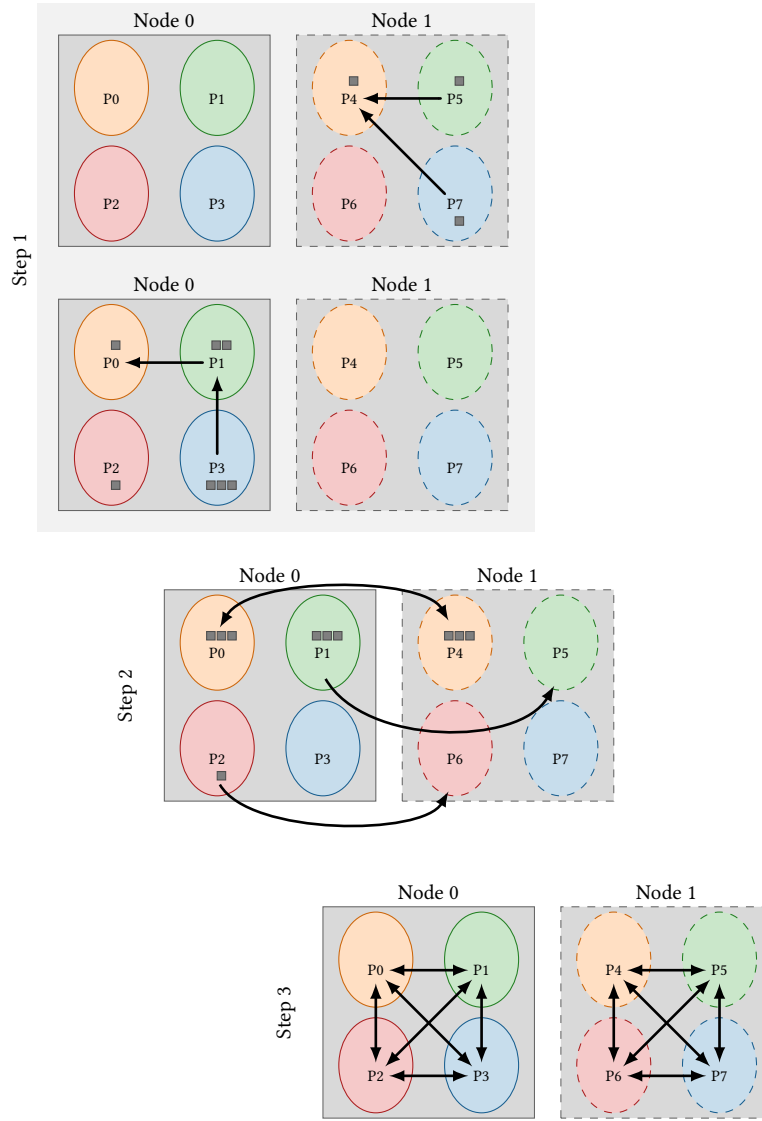

The proposed process, excluding reducing the global
communication strategy to 3-step or 2-step communication, is summarized in~\cref{fig:optstep}.
This figure outlines the nodal optimal process of
communicating data between two nodes, each
with four local processes.
\begin{description}
\item[Step 1:] Each node conglomerates small messages to be sent off-node
  while retaining larger messages. Messages are assigned to processes in
  descending order of size to the first available process on node.
  This is done for every node simultaneously.
\item[Step 2:]  Buffers prepared in step 1 are sent to their destination node,
  specifically to the paired process on that node with the same local rank.
  P0 exchanges data with P4, while P1 sends data to P5, and P2 sends
  data to P6.
\item[Step 3:]
  All processes on each individual node redistribute their received data to the
  correct destination processes on-node.
  In this step, all communication is local.
\end{description}

\begin{figure}[!ht]
  \centering
\includegraphics[width=1.0\textwidth]{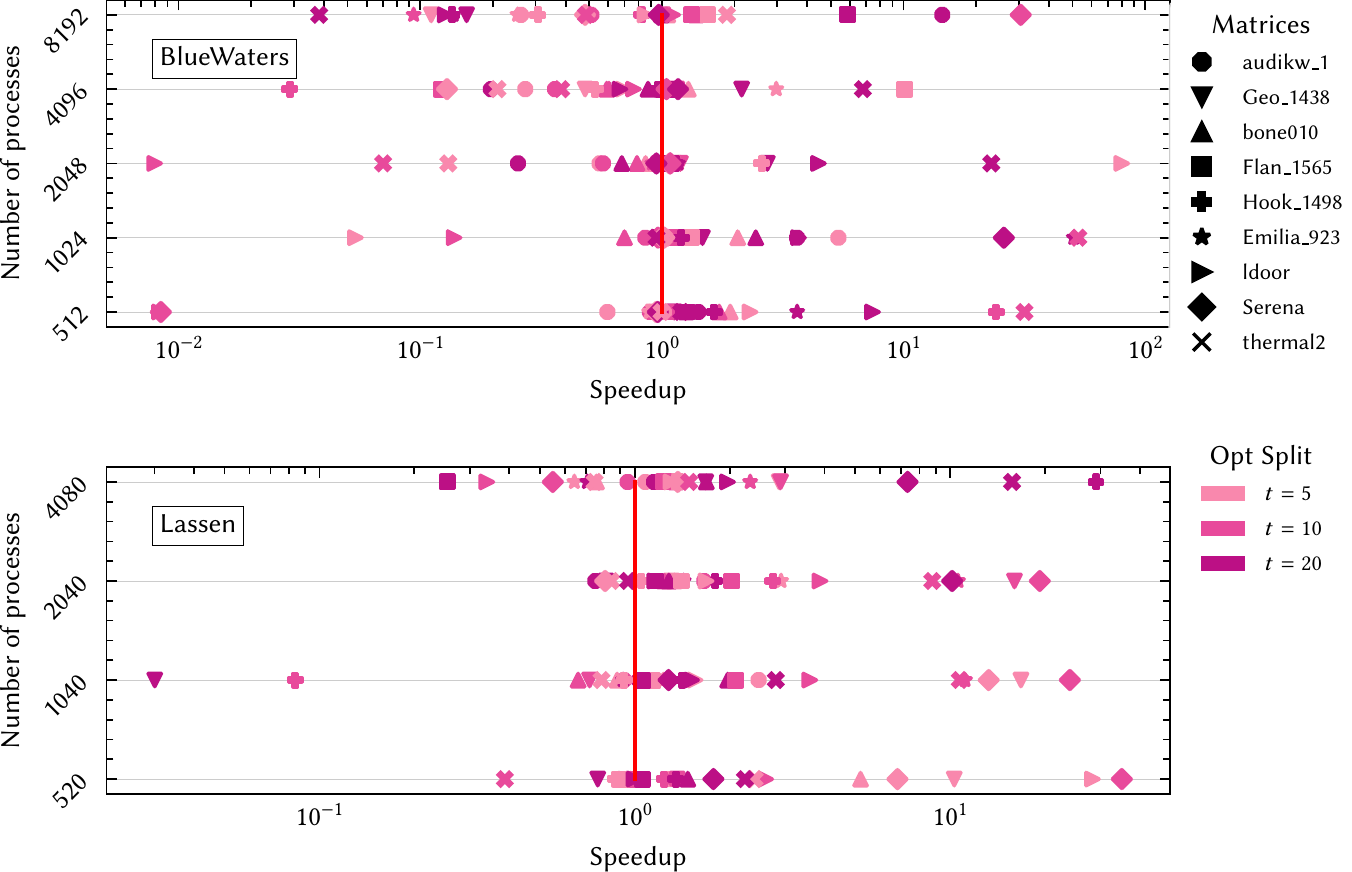}
\caption{Speedup of optimal communication over standard communication
(without reducing to a global communication strategy) in the SpMBV for various matrices
from the SuiteSparse matrix collection on Blue Waters and Lassen.
The red line marks 1, or no speedup.}\label{fig:bw_lassen_spmbv_opt}
\end{figure}

The resulting speedups for the two systems Blue Waters and Lassen are presented in
\cref{fig:bw_lassen_spmbv_opt}. Here, the message size cutoff being
used is the message size cutoff before the MPI implementation switches to the rendezvous protocol. For sending large messages between processes, the rendezvous protocol communicates an envelope first, then
the remaining data is communicated after the receiving process allocates buffer space.
It is necessary to use this protocol for large messages, but there is a slight slowdown over sending messages via the short and eager protocols which either include the data being
sent as part of the envelope (short) or eagerly send the data if it does not fit into the
envelope (eager).
This cutoff is chosen because rendezvous is the slowest protocol and because the switch
to the rendezvous protocol is approximately the crossover point when on-node messages
become slower than network messages on Lassen (around $10^4$ Bytes in \cref{fig:comm_split}.)

Using this cutoff point, the method sees speedup for \textit{some} test matrices and performance degradation
for most on Blue Waters,
which is consistent with the minimal speedup seen by splitting large messages
across multiple processes in \cref{fig:node_pong}.
Additionally, it is likely that network contention is playing a large role in the
Blue Waters results as the messages sizes become large based on the findings
in~\cite{Bienz_model}, and due to
each node determining independently how to send its own data
without consideration of the size or number of messages injected by other nodes.

\begin{figure}[!ht]
  \centering
\includegraphics[width=1.0\textwidth]{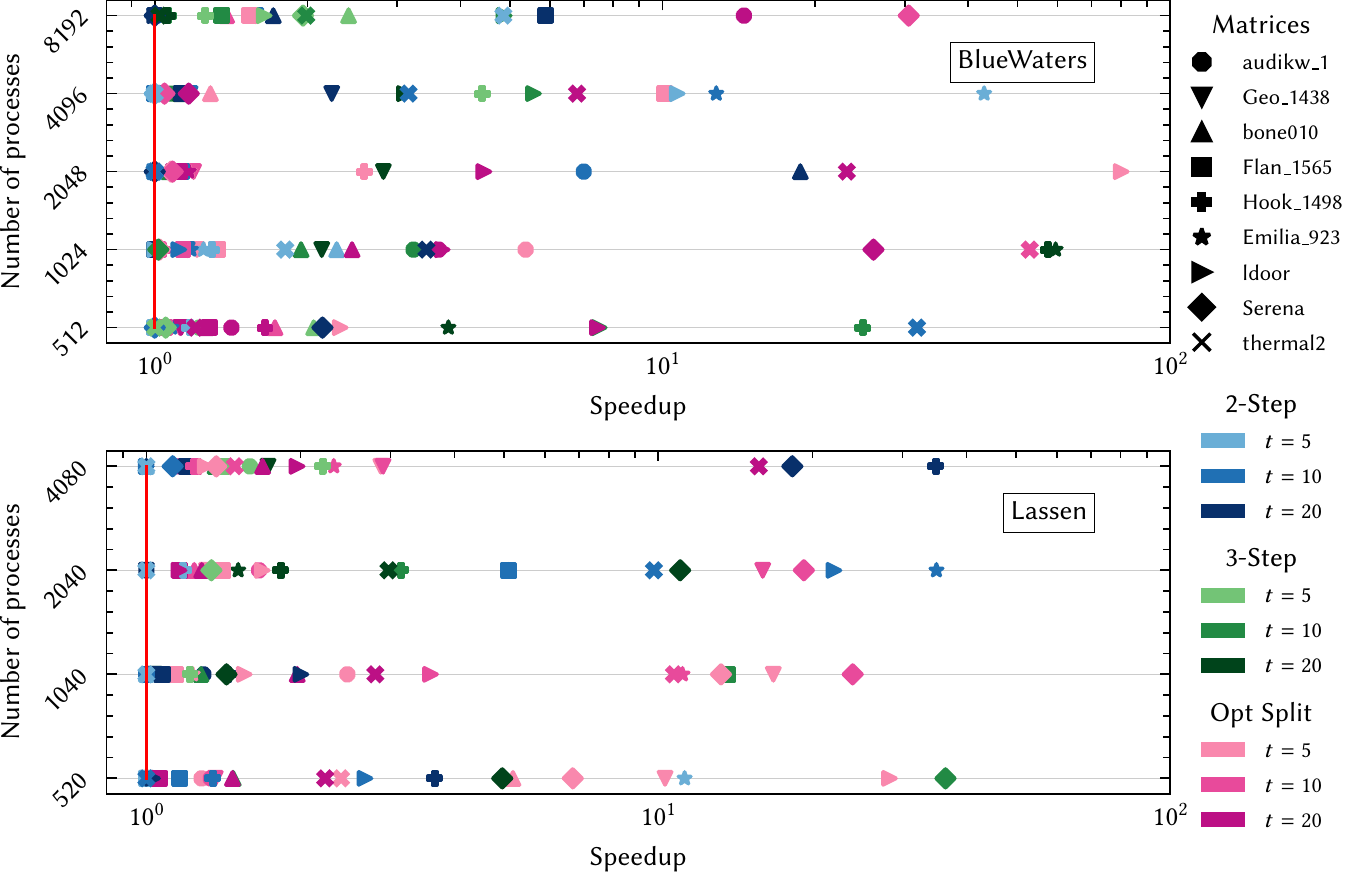}
\caption{Speedup of tuned communication over standard communication
in the SpMBV for various matrices from the SuiteSparse matrix collection on Blue Waters and Lassen.
The red line marks 1.0, or no speedup.
}\label{fig:bw_lassen_spmbv_min}
\end{figure}

The nodal optimal communication performs better on Lassen than Blue Waters and
aligns with expectations based on the combination
of using 3-step for nodes with small inter-node messages to inject into the network where
on-node communication is faster than network communication (\cref{fig:comm_split})
and splitting large messages where the benefits are much greater (\cref{fig:node_pong}). While Blue Waters achieves higher speedups in some
cases than Lassen, both systems see speedups as large as 60x.
These results only present part of the overall communication picture, however.
There are still cases where the global communication strategy should be reduced
to 3-step communication, 2-step communication, or standard communication.
This differs based on the two machines and the specific test matrix, but
tuning between the techniques results in the most optimal communication strategy.
Tuning comes at the minimal cost of performing four different SpMBVs during
setup of the SpMBV communicator.
Speedups over standard communication when using tuning to
use the fastest communication technique
are presented in \cref{fig:bw_lassen_spmbv_min}.
In the top plot of \cref{fig:bw_lassen_spmbv_min},
Blue Waters benefits in 33\% of the cases from including
the nodal optimal multi-step communication strategy. These results are expected based on
\cref{fig:bw_lassen_spmbv_opt} where the benefits of nodal optimal multi-step communication
were less than ideal. In fact, for 20\% of the cases on Blue Waters, standard
communication is the most performant, consistent with matrices of low density.
The matrices \texttt{Geo\_1438} and \texttt{ldoor}, which have the smallest density of the test
matrices (\cref{tab:matrices}), see minimal
benefits from the multi-step communication techniques
as the number of processes is scaled up
due to the minimal amount of data being communicated.

We expected to see 2-step and nodal optimal communication perform the best
on Lassen due to the faster inter-node communication for smaller sized messages
(\cref{fig:comm_split}) and the benefits of splitting messages across many
processes on node (\cref{fig:node_pong}). These expectations are consistent with the
results presented in the bottom plot of \cref{fig:bw_lassen_spmbv_min};
most test matrices saw the best SpMBV performance with nodal optimal communication
(44\% of the cases). The remainder of the test cases were divided almost equally between
2-step, 3-step, and standard communication for which technique was the most performant
(approximately 18\% of the cases, each).

\cref{tab:ecg_percents} shows the benefits of using the tuned
point-to-point communication over the standard communication in a
single iteration of ECG for \cref{ex:mfem}.
Tuned communication reduces the percentage of time spent in
point-to-point communication independent of system,
though the performance benefits are typically best when
more data is being communicated ($t=20$ in \cref{tab:ecg_percents_bw}
and \cref{tab:ecg_percents_lassen}.)
For Blue Waters, the new communication technique results in
point-to-point communication
taking $20\% - 40\%$ less of the total time compared to the
percentage of time when using standard communication
(corresponding to the blue highlighted values in \cref{tab:ecg_percents_bw}).
Performance benefits are much greater on Lassen where the
tuned communication results in a decrease of more than $40\%$
of the total iteration time compared to an iteration time
with standard communication in most cases.
\begin{table}
\begin{subtable}{\textwidth}
\centering
  \begin{tabular}{c|cc|cc|cc|cc|cc|cc}
    \toprule
    \multicolumn{1}{c}{procs $\rightarrow$}    &
    \multicolumn{2}{c}{256} &
    \multicolumn{2}{c}{512} &
    \multicolumn{2}{c}{1024} &
    \multicolumn{2}{c}{2048} &
    \multicolumn{2}{c}{4096} &
    \multicolumn{2}{c}{8192}\\
    \multicolumn{1}{c}{$t$} &
    \scriptsize{m-s} & \multicolumn{1}{c}{\scriptsize{std}} &
    \scriptsize{m-s} & \multicolumn{1}{c}{\scriptsize{std}} &
    \scriptsize{m-s} & \multicolumn{1}{c}{\scriptsize{std}} &
    \scriptsize{m-s} & \multicolumn{1}{c}{\scriptsize{std}} &
    \scriptsize{m-s} & \multicolumn{1}{c}{\scriptsize{std}} &
    \scriptsize{m-s} & \multicolumn{1}{c}{\scriptsize{std}}\\
    \midrule
    5  & 54.7 & 55.6 & \hlg{34.1} & {70.0} & 66.2 & 70.2 & \hlb{70.6} & {75.5} & \hlb{70.2} & {78.9} & \hlg{41.7} & {76.6} \\
    10 & \hlb{31.8} & {40.0} & \hlb{39.3} & {56.2} & \hlg{34.6} & {65.2} & \hlb{48.4} & {67.5} & \hlb{59.7} & {69.1} & \hlg{32.2} & {68.7} \\
    15 & \hlg{9.3} & {37.6} & \hlg{18.2} & {40.4} & \hlg{30.4} & {66.7} & \hlg{31.9} & {67.0} & \hlg{38.4} & {58.5} & \hlg{26.0} & {63.8} \\
    20 & \hlb{7.5} & {24.5} & \hlg{15.0} & {38.3} & \hlg{26.3} & {62.3} & 46.0 & 47.6 & \hlb{30.8} & {45.5} & \hlg{29.7} & {53.6} \\
    \bottomrule
  \end{tabular}
  \caption{Blue Waters}\label{tab:ecg_percents_bw}
\end{subtable}

\bigskip
\begin{subtable}{\textwidth}
\centering
  \begin{tabular}{c|cc|cc|cc|cc|cc}
    \toprule
    \multicolumn{1}{c}{procs $\rightarrow$}    &
    \multicolumn{2}{c}{240} &
    \multicolumn{2}{c}{520} &
    \multicolumn{2}{c}{1040} &
    \multicolumn{2}{c}{2040} &
    \multicolumn{2}{c}{4080} \\
    \multicolumn{1}{c}{$t$} &
    \scriptsize{m-s} & \multicolumn{1}{c}{\scriptsize{std}} &
    \scriptsize{m-s} & \multicolumn{1}{c}{\scriptsize{std}} &
    \scriptsize{m-s} & \multicolumn{1}{c}{\scriptsize{std}} &
    \scriptsize{m-s} & \multicolumn{1}{c}{\scriptsize{std}} &
    \scriptsize{m-s} & \multicolumn{1}{c}{\scriptsize{std}} \\
    \midrule
    5 & \hlg{49.4} & {84.3} & \hly{43.6} & {88.0} & \hlg{66.6} & {88.4} & \hlb{82.2} & {87.2} &  \hlb{82.7} & {93.1} \\
    10 & \hly{26.8} & {87.2} & \hly{33.7} & {84.9} & \hlg{52.1} & {84.7} & \hlg{58.4} & {89.3} & \hlb{78.3} & {91.3} \\
    15 & \hly{9.9} & {81.2} & \hly{26.4} & {85.7} & \hly{33.6} & {86.1} & \hlg{55.3} & {90.0} & \hlg{61.2} & {89.2} \\
    20 & \hly{12.7} & {74.2} & \hly{17.8} & {79.9} & \hly{29.6} & {81.7} & \hly{39.0} & {80.7} & \hlg{57.2} & {79.5} \\
    \bottomrule
  \end{tabular}
  \caption{Lassen}\label{tab:ecg_percents_lassen}
\end{subtable}
\caption{Measured percentage of time spent in point-to-point communication for the tuned
    multistep (``m-s'') compared against the measured percentage of time spent
    in point-to-point communication for standard (``std'') in a single iteration of ECG
    for \cref{ex:mfem} with varying $t$ and processor counts on Blue Waters and Lassen.
    Blue values correspond to values for which the percentage of time
    \hlc[hlblue]{decreased by $5\% - 20\%$}, green values are a
    \hlc[hlgreen]{decrease of $20\% - 40\%$}, and yellow values
    \hlc[hlyellow]{decreased by more than 40\%} from the original percentage.
  }\label{tab:ecg_percents}
\end{table}

\section{Conclusions}\label{sec:conclusion}

The enlarged conjugate gradient method (ECG) is an efficient method for solving large
systems of equations designed to reduce the collective communication bottlenecks of the
classical conjugate gradient method (CG.) Within ECG, block vector updates replace the
single vector updates of CG, thereby reducing the overall number of iterations required for
convergence and hence the overall amount of collective communication.

In this paper, we performed a performance study and analysis of the effects of block
vectors on the balance of collective communication, point-to-point communication, and
computation within the iterations of ECG\@. We noted the increased volume of data communicated
and its disproportionate affects on the performance of the point-to-point communication;
the communication bottleneck of ECG shifted to be the point-to-point communication within
the sparse matrix-block vector kernel (SpMBV).
To address the new SpMBV bottleneck, we designed an optimal multi-step communication
technique that builds on existing node-aware communication techniques and improves them
for the emerging supercomputer architectures with greater numbers of processes per node
and faster inter-node networks.

Overall, this paper provides a comprehensive study of the performance of ECG in
a distributed parallel environment, and introduces a novel point-to-point multi-step
communication technique that provides consistent speedup over standard communication techniques
independent of machine.
Future work includes profiling and improving the performance of
ECG on hybrid supercomputer architectures with computation
offloaded to graphics processing units.

The software used to generate the results in this
paper is freely available in RAPtor~\cite{raptor}.

\section*{Acknowledgments}\label{sec:acknowledgements}
This material is based in part upon work supported by the Department of Energy,
National Nuclear Security Administration, under Award Number \textit{DE-NA0003963} and \textit{DE-NA0003966}.

This research is part of the Blue Waters sustained-petascale computing
project, which is supported by the National Science Foundation (awards OCI0725070 and ACI-1238993) and the state of Illinois. Blue Waters is a joint
effort of the University of Illinois at Urbana-Champaign and its National
Center for Supercomputing Applications. This material is based in part upon
work supported by the Department of Energy, National Nuclear Security
Administration, under Award Number de-na0002374.

\bibliographystyle{ACM-Reference-Format}
\bibliography{references}


\begin{thebibliography}{23}


\ifx \showCODEN    \undefined \def \showCODEN     #1{\unskip}     \fi
\ifx \showDOI      \undefined \def \showDOI       #1{#1}\fi
\ifx \showISBNx    \undefined \def \showISBNx     #1{\unskip}     \fi
\ifx \showISBNxiii \undefined \def \showISBNxiii  #1{\unskip}     \fi
\ifx \showISSN     \undefined \def \showISSN      #1{\unskip}     \fi
\ifx \showLCCN     \undefined \def \showLCCN      #1{\unskip}     \fi
\ifx \shownote     \undefined \def \shownote      #1{#1}          \fi
\ifx \showarticletitle \undefined \def \showarticletitle #1{#1}   \fi
\ifx \showURL      \undefined \def \showURL       {\relax}        \fi
\providecommand\bibfield[2]{#2}
\providecommand\bibinfo[2]{#2}
\providecommand\natexlab[1]{#1}
\providecommand\showeprint[2][]{arXiv:#2}

\bibitem[\protect\citeauthoryear{Alexandrov, Ionescu, Schauser, and
  Scheiman}{Alexandrov et~al\mbox{.}}{1997}]%
        {logGP}
\bibfield{author}{\bibinfo{person}{Albert Alexandrov},
  \bibinfo{person}{Mihai~F. Ionescu}, \bibinfo{person}{Klaus~E. Schauser},
  {and} \bibinfo{person}{Chris Scheiman}.} \bibinfo{year}{1997}\natexlab{}.
\newblock \showarticletitle{LogGP: Incorporating Long Messages into the LogP
  Model for Parallel Computation}.
\newblock \bibinfo{journal}{\emph{J. Parallel and Distrib. Comput.}}
  \bibinfo{volume}{44}, \bibinfo{number}{1} (\bibinfo{year}{1997}),
  \bibinfo{pages}{71 -- 79}.
\newblock
\showISSN{0743-7315}
\urldef\tempurl%
\url{https://doi.org/10.1006/jpdc.1997.1346}
\showDOI{\tempurl}


\bibitem[\protect\citeauthoryear{Anderson, Andrej, Barker, Bramwell, Camier,
  Cerveny, Dobrev, Dudouit, Fisher, Kolev, Pazner, Stowell, Tomov, Akkerman,
  Dahm, Medina, and Zampini}{Anderson et~al\mbox{.}}{2020}]%
        {mfem-library}
\bibfield{author}{\bibinfo{person}{Robert Anderson}, \bibinfo{person}{Julian
  Andrej}, \bibinfo{person}{Andrew Barker}, \bibinfo{person}{Jamie Bramwell},
  \bibinfo{person}{Jean-Sylvain Camier}, \bibinfo{person}{Jakub Cerveny},
  \bibinfo{person}{Veselin Dobrev}, \bibinfo{person}{Yohann Dudouit},
  \bibinfo{person}{Aaron Fisher}, \bibinfo{person}{Tzanio Kolev},
  \bibinfo{person}{Will Pazner}, \bibinfo{person}{Mark Stowell},
  \bibinfo{person}{Vladimir Tomov}, \bibinfo{person}{Ido Akkerman},
  \bibinfo{person}{Johann Dahm}, \bibinfo{person}{David Medina}, {and}
  \bibinfo{person}{Stefano Zampini}.} \bibinfo{year}{2020}\natexlab{}.
\newblock \showarticletitle{{MFEM}: A modular finite element methods library}.
\newblock \bibinfo{journal}{\emph{Computers \& Mathematics with Applications}}
  (\bibinfo{year}{2020}).
\newblock
\showISSN{0898-1221}
\urldef\tempurl%
\url{https://doi.org/10.1016/j.camwa.2020.06.009}
\showDOI{\tempurl}


\bibitem[\protect\citeauthoryear{Bienz, Gropp, and Olson}{Bienz
  et~al\mbox{.}}{2019}]%
        {Bienz_napspmv}
\bibfield{author}{\bibinfo{person}{Amanda Bienz}, \bibinfo{person}{William
  Gropp}, {and} \bibinfo{person}{Luke Olson}.} \bibinfo{year}{2019}\natexlab{}.
\newblock \showarticletitle{Node aware sparse matrix–vector multiplication}.
\newblock \bibinfo{journal}{\emph{J. Parallel and Distrib. Comput.}}
  \bibinfo{volume}{130} (\bibinfo{date}{08} \bibinfo{year}{2019}),
  \bibinfo{pages}{166--178}.
\newblock
\urldef\tempurl%
\url{https://doi.org/10.1016/j.jpdc.2019.03.016}
\showDOI{\tempurl}


\bibitem[\protect\citeauthoryear{Bienz, Gropp, and Olson}{Bienz
  et~al\mbox{.}}{2018}]%
        {Bienz_model}
\bibfield{author}{\bibinfo{person}{Amanda Bienz}, \bibinfo{person}{William~D.
  Gropp}, {and} \bibinfo{person}{Luke~N. Olson}.}
  \bibinfo{year}{2018}\natexlab{}.
\newblock \showarticletitle{Improving Performance Models for Irregular
  Point-to-Point Communication}.
\newblock \bibinfo{journal}{\emph{CoRR}}  \bibinfo{volume}{abs/1806.02030}
  (\bibinfo{year}{2018}).
\newblock
\urldef\tempurl%
\url{http://arxiv.org/abs/1806.02030}
\showURL{%
\tempurl}


\bibitem[\protect\citeauthoryear{Bienz, Gropp, and Olson}{Bienz
  et~al\mbox{.}}{2020}]%
        {Bienz_napamg}
\bibfield{author}{\bibinfo{person}{Amanda Bienz}, \bibinfo{person}{William~D
  Gropp}, {and} \bibinfo{person}{Luke~N Olson}.}
  \bibinfo{year}{2020}\natexlab{}.
\newblock \showarticletitle{Reducing communication in algebraic multigrid with
  multi-step node aware communication}.
\newblock \bibinfo{journal}{\emph{The International Journal of High Performance
  Computing Applications}} \bibinfo{volume}{34}, \bibinfo{number}{5}
  (\bibinfo{date}{June} \bibinfo{year}{2020}), \bibinfo{pages}{547--561}.
\newblock
\showISSN{1094-3420, 1741-2846}
\urldef\tempurl%
\url{https://doi.org/10.1177/1094342020925535}
\showDOI{\tempurl}


\bibitem[\protect\citeauthoryear{Bienz and Olson}{Bienz and Olson}{2017}]%
        {raptor}
\bibfield{author}{\bibinfo{person}{Amanda Bienz} {and} \bibinfo{person}{Luke~N.
  Olson}.} \bibinfo{year}{2017}\natexlab{}.
\newblock \bibinfo{title}{{RAPtor}: parallel algebraic multigrid v0.1}.
\newblock
  \bibinfo{howpublished}{\url{https://github.com/raptor-library/raptor}}.
\newblock
\newblock
\shownote{Release 0.1.}


\bibitem[\protect\citeauthoryear{Bode, Butler, Dunning, Hoefler, Kramer, Gropp,
  and Hwu}{Bode et~al\mbox{.}}{2013}]%
        {BW1}
\bibfield{author}{\bibinfo{person}{Brett Bode}, \bibinfo{person}{Michelle
  Butler}, \bibinfo{person}{Thom Dunning}, \bibinfo{person}{Torsten Hoefler},
  \bibinfo{person}{William Kramer}, \bibinfo{person}{William Gropp}, {and}
  \bibinfo{person}{Wen-mei Hwu}.} \bibinfo{year}{2013}\natexlab{}.
\newblock \showarticletitle{The {B}lue {W}aters Super-System for
  Super-Science}.
\newblock In \bibinfo{booktitle}{\emph{Contemporary High Performance
  Computing}}. \bibinfo{publisher}{Chapman and Hall/CRC},
  \bibinfo{pages}{339--366}.
\newblock
\showISBNx{978-1-4665-6834-1}
\urldef\tempurl%
\url{https://www.taylorfrancis.com/books/e/9781466568358}
\showURL{%
\tempurl}


\bibitem[\protect\citeauthoryear{Carson, Knight, and Demmel}{Carson
  et~al\mbox{.}}{2013}]%
        {avoidcomm_lanzcos}
\bibfield{author}{\bibinfo{person}{Erin Carson}, \bibinfo{person}{Nicholas
  Knight}, {and} \bibinfo{person}{James Demmel}.}
  \bibinfo{year}{2013}\natexlab{}.
\newblock \showarticletitle{Avoiding Communication in Nonsymmetric
  Lanczos-Based Krylov Subspace Methods}.
\newblock \bibinfo{journal}{\emph{SIAM J. Sci. Comput.}} \bibinfo{volume}{35},
  \bibinfo{number}{5} (\bibinfo{date}{Jan.} \bibinfo{year}{2013}),
  \bibinfo{pages}{S42--S61}.
\newblock
\showISSN{1064-8275, 1095-7197}
\urldef\tempurl%
\url{https://doi.org/10.1137/120881191}
\showDOI{\tempurl}
\showeprint{https://doi.org/10.1137/120881191}


\bibitem[\protect\citeauthoryear{Chronopoulos and Gear}{Chronopoulos and
  Gear}{1989}]%
        {s_stepIter}
\bibfield{author}{\bibinfo{person}{A.T. Chronopoulos} {and}
  \bibinfo{person}{C.W. Gear}.} \bibinfo{year}{1989}\natexlab{}.
\newblock \showarticletitle{s-step iterative methods for symmetric linear
  systems}.
\newblock \bibinfo{journal}{\emph{J. Comput. Appl. Math.}}
  \bibinfo{volume}{25}, \bibinfo{number}{2} (\bibinfo{date}{Feb.}
  \bibinfo{year}{1989}), \bibinfo{pages}{153--168}.
\newblock
\showISSN{0377-0427}
\urldef\tempurl%
\url{https://doi.org/10.1016/0377-0427(89)90045-9}
\showDOI{\tempurl}


\bibitem[\protect\citeauthoryear{Cools and Vanroose}{Cools and
  Vanroose}{2017}]%
        {pipelined}
\bibfield{author}{\bibinfo{person}{S. Cools} {and} \bibinfo{person}{W.
  Vanroose}.} \bibinfo{year}{2017}\natexlab{}.
\newblock \showarticletitle{The communication-hiding pipelined {BiCGstab}
  method for the parallel solution of large unsymmetric linear systems}.
\newblock \bibinfo{journal}{\emph{Parallel Comput.}}  \bibinfo{volume}{65}
  (\bibinfo{date}{July} \bibinfo{year}{2017}), \bibinfo{pages}{1--20}.
\newblock
\showISSN{0167-8191}
\urldef\tempurl%
\url{https://doi.org/10.1016/j.parco.2017.04.005}
\showDOI{\tempurl}


\bibitem[\protect\citeauthoryear{Culler, Karp, Patterson, Sahay, Schauser,
  Santos, Subramonian, and von Eicken}{Culler et~al\mbox{.}}{1993}]%
        {logP}
\bibfield{author}{\bibinfo{person}{David Culler}, \bibinfo{person}{Richard
  Karp}, \bibinfo{person}{David Patterson}, \bibinfo{person}{Abhijit Sahay},
  \bibinfo{person}{Klaus~Erik Schauser}, \bibinfo{person}{Eunice Santos},
  \bibinfo{person}{Ramesh Subramonian}, {and} \bibinfo{person}{Thorsten von
  Eicken}.} \bibinfo{year}{1993}\natexlab{}.
\newblock \showarticletitle{LogP: Towards a Realistic Model of Parallel
  Computation}.
\newblock \bibinfo{journal}{\emph{SIGPLAN Not.}} \bibinfo{volume}{28},
  \bibinfo{number}{7} (\bibinfo{date}{July} \bibinfo{year}{1993}),
  \bibinfo{pages}{1–12}.
\newblock
\showISSN{0362-1340}
\urldef\tempurl%
\url{https://doi.org/10.1145/173284.155333}
\showDOI{\tempurl}


\bibitem[\protect\citeauthoryear{Davis and Hu}{Davis and Hu}{2011}]%
        {ufl_matrices}
\bibfield{author}{\bibinfo{person}{Timothy~A. Davis} {and}
  \bibinfo{person}{Yifan Hu}.} \bibinfo{year}{2011}\natexlab{}.
\newblock \showarticletitle{The University of Florida Sparse Matrix
  Collection}.
\newblock \bibinfo{journal}{\emph{ACM Trans. Math. Softw.}}
  \bibinfo{volume}{38}, \bibinfo{number}{1}, Article \bibinfo{articleno}{1}
  (\bibinfo{date}{Dec.} \bibinfo{year}{2011}), \bibinfo{numpages}{25}~pages.
\newblock
\showISSN{0098-3500}
\urldef\tempurl%
\url{https://doi.org/10.1145/2049662.2049663}
\showDOI{\tempurl}


\bibitem[\protect\citeauthoryear{Ghysels and Vanroose}{Ghysels and
  Vanroose}{2014}]%
        {ghysels2014hiding}
\bibfield{author}{\bibinfo{person}{P. Ghysels} {and} \bibinfo{person}{W.
  Vanroose}.} \bibinfo{year}{2014}\natexlab{}.
\newblock \showarticletitle{Hiding global synchronization latency in the
  preconditioned Conjugate Gradient algorithm}.
\newblock \bibinfo{journal}{\emph{Parallel Comput.}} \bibinfo{volume}{40},
  \bibinfo{number}{7} (\bibinfo{date}{July} \bibinfo{year}{2014}),
  \bibinfo{pages}{224--238}.
\newblock
\showISSN{0167-8191}
\urldef\tempurl%
\url{https://doi.org/10.1016/j.parco.2013.06.001}
\showDOI{\tempurl}


\bibitem[\protect\citeauthoryear{Grigori, Moufawad, and Nataf}{Grigori
  et~al\mbox{.}}{2016}]%
        {enlarged}
\bibfield{author}{\bibinfo{person}{Laura Grigori}, \bibinfo{person}{Sophie
  Moufawad}, {and} \bibinfo{person}{Frederic Nataf}.}
  \bibinfo{year}{2016}\natexlab{}.
\newblock \showarticletitle{Enlarged Krylov Subspace Conjugate Gradient Methods
  for Reducing Communication}.
\newblock \bibinfo{journal}{\emph{SIAM J. Matrix Anal. \& Appl.}}
  \bibinfo{volume}{37}, \bibinfo{number}{2} (\bibinfo{date}{Jan.}
  \bibinfo{year}{2016}), \bibinfo{pages}{744--773}.
\newblock
\showISSN{0895-4798, 1095-7162}
\urldef\tempurl%
\url{https://doi.org/10.1137/140989492}
\showDOI{\tempurl}
\showeprint{https://doi.org/10.1137/140989492}


\bibitem[\protect\citeauthoryear{Grigori and Tissot}{Grigori and
  Tissot}{2019}]%
        {scalable_enlarged}
\bibfield{author}{\bibinfo{person}{Laura Grigori} {and}
  \bibinfo{person}{Olivier Tissot}.} \bibinfo{year}{2019}\natexlab{}.
\newblock \showarticletitle{Scalable Linear Solvers Based on Enlarged Krylov
  Subspaces with Dynamic Reduction of Search Directions}.
\newblock \bibinfo{journal}{\emph{SIAM J. Sci. Comput.}} \bibinfo{volume}{41},
  \bibinfo{number}{5} (\bibinfo{date}{Jan.} \bibinfo{year}{2019}),
  \bibinfo{pages}{C522--C547}.
\newblock
\showISSN{1064-8275, 1095-7197}
\urldef\tempurl%
\url{https://doi.org/10.1137/18m1196285}
\showDOI{\tempurl}
\showeprint{https://doi.org/10.1137/18M1196285}


\bibitem[\protect\citeauthoryear{Gropp, Olson, and Samfass}{Gropp
  et~al\mbox{.}}{2016}]%
        {maxrate_model}
\bibfield{author}{\bibinfo{person}{William Gropp}, \bibinfo{person}{Luke~N.
  Olson}, {and} \bibinfo{person}{Philipp Samfass}.}
  \bibinfo{year}{2016}\natexlab{}.
\newblock \showarticletitle{Modeling {MPI} Communication Performance on {SMP}
  Nodes}. In \bibinfo{booktitle}{\emph{Proceedings of the 23rd European MPI
  Users' Group Meeting on - EuroMPI 2016}} \emph{(\bibinfo{series}{EuroMPI
  2016})}. \bibinfo{publisher}{ACM Press}, \bibinfo{address}{New York, NY,
  USA}, \bibinfo{pages}{41--50}.
\newblock
\showISBNx{9781450342346}
\urldef\tempurl%
\url{https://doi.org/10.1145/2966884.2966919}
\showDOI{\tempurl}


\bibitem[\protect\citeauthoryear{{Hanson}}{{Hanson}}{2020}]%
        {lassen}
\bibfield{author}{\bibinfo{person}{W.~A. {Hanson}}.}
  \bibinfo{year}{2020}\natexlab{}.
\newblock \showarticletitle{The {CORAL} supercomputer systems}.
\newblock \bibinfo{journal}{\emph{IBM Journal of Research and Development}}
  \bibinfo{volume}{64}, \bibinfo{number}{3/4} (\bibinfo{year}{2020}),
  \bibinfo{pages}{1:1--1:10}.
\newblock


\bibitem[\protect\citeauthoryear{Kramer, Butler, Bauer, Chadalavada, and
  Mendes}{Kramer et~al\mbox{.}}{2015}]%
        {BW2}
\bibfield{author}{\bibinfo{person}{William Kramer}, \bibinfo{person}{Michelle
  Butler}, \bibinfo{person}{Gregory Bauer}, \bibinfo{person}{Kalyana
  Chadalavada}, {and} \bibinfo{person}{Celso Mendes}.}
  \bibinfo{year}{2015}\natexlab{}.
\newblock \showarticletitle{{Blue Waters Parallel I/O Storage Sub-system}}.
\newblock In \bibinfo{booktitle}{\emph{High Performance Parallel I/O}},
  \bibfield{editor}{\bibinfo{person}{Prabhat} {and} \bibinfo{person}{Quincey
  Koziol}} (Eds.). \bibinfo{publisher}{CRC Publications, Taylor and Francis
  Group}, \bibinfo{pages}{17--32}.
\newblock
\showISBNx{978-1-4665-8234-7}


\bibitem[\protect\citeauthoryear{M.}{M.}{2010}]%
        {HoemmenThesis}
\bibfield{author}{\bibinfo{person}{Hoemmen M.}}
  \bibinfo{year}{2010}\natexlab{}.
\newblock \emph{\bibinfo{title}{Communication-Avoiding Krylov Subspace
  Methods}}.
\newblock \bibinfo{thesistype}{Ph.D. Dissertation}. \bibinfo{school}{University
  of California, Berkeley}.
\newblock


\bibitem[\protect\citeauthoryear{McInnes, Smith, Zhang, and Mills}{McInnes
  et~al\mbox{.}}{2014}]%
        {mcinnes2014hierarchical}
\bibfield{author}{\bibinfo{person}{Lois~Curfman McInnes},
  \bibinfo{person}{Barry Smith}, \bibinfo{person}{Hong Zhang}, {and}
  \bibinfo{person}{Richard~Tran Mills}.} \bibinfo{year}{2014}\natexlab{}.
\newblock \showarticletitle{Hierarchical Krylov and nested Krylov methods for
  extreme-scale computing}.
\newblock \bibinfo{journal}{\emph{Parallel Comput.}} \bibinfo{volume}{40},
  \bibinfo{number}{1} (\bibinfo{date}{Jan.} \bibinfo{year}{2014}),
  \bibinfo{pages}{17--31}.
\newblock
\showISSN{0167-8191}
\urldef\tempurl%
\url{https://doi.org/10.1016/j.parco.2013.10.001}
\showDOI{\tempurl}


\bibitem[\protect\citeauthoryear{Mohiyuddin, Hoemmen, Demmel, and
  Yelick}{Mohiyuddin et~al\mbox{.}}{2009}]%
        {mohiyuddin2009minimizing}
\bibfield{author}{\bibinfo{person}{Marghoob Mohiyuddin}, \bibinfo{person}{Mark
  Hoemmen}, \bibinfo{person}{James Demmel}, {and} \bibinfo{person}{Katherine
  Yelick}.} \bibinfo{year}{2009}\natexlab{}.
\newblock \showarticletitle{Minimizing Communication in Sparse Matrix Solvers}.
  In \bibinfo{booktitle}{\emph{Proceedings of the Conference on High
  Performance Computing Networking, Storage and Analysis}}
  \emph{(\bibinfo{series}{SC '09})}. \bibinfo{publisher}{Association for
  Computing Machinery}, \bibinfo{address}{New York, NY, USA}, Article
  \bibinfo{articleno}{36}, \bibinfo{numpages}{12}~pages.
\newblock
\showISBNx{9781605587448}
\urldef\tempurl%
\url{https://doi.org/10.1145/1654059.1654096}
\showDOI{\tempurl}


\bibitem[\protect\citeauthoryear{Moufawad}{Moufawad}{2020}]%
        {sstep_enlarged}
\bibfield{author}{\bibinfo{person}{Sophie~M. Moufawad}.}
  \bibinfo{year}{2020}\natexlab{}.
\newblock \showarticletitle{s-Step Enlarged Krylov Subspace Conjugate Gradient
  Methods}.
\newblock \bibinfo{journal}{\emph{SIAM J. Sci. Comput.}} \bibinfo{volume}{42},
  \bibinfo{number}{1} (\bibinfo{date}{Jan.} \bibinfo{year}{2020}),
  \bibinfo{pages}{A187--A219}.
\newblock
\showISSN{1064-8275, 1095-7197}
\urldef\tempurl%
\url{https://doi.org/10.1137/18m1182528}
\showDOI{\tempurl}
\showeprint{https://doi.org/10.1137/18M1182528}


\bibitem[\protect\citeauthoryear{O'Leary}{O'Leary}{1980}]%
        {block}
\bibfield{author}{\bibinfo{person}{Dianne~P. O'Leary}.}
  \bibinfo{year}{1980}\natexlab{}.
\newblock \showarticletitle{The block conjugate gradient algorithm and related
  methods}.
\newblock \bibinfo{journal}{\emph{Linear Algebra Appl.}}  \bibinfo{volume}{29}
  (\bibinfo{date}{Feb.} \bibinfo{year}{1980}), \bibinfo{pages}{293--322}.
\newblock
\showISSN{0024-3795}
\urldef\tempurl%
\url{https://doi.org/10.1016/0024-3795(80)90247-5}
\showDOI{\tempurl}


\end{thebibliography}

\end{document}